\documentclass[aps,pra,twocolumn,showpacs,superscriptaddress]{revtex4} 
\usepackage{graphicx}
\usepackage{amsmath}
\usepackage{amssymb}
\usepackage{epstopdf}
\usepackage{amsfonts}
\usepackage{dcolumn}
\usepackage{bigints}
\usepackage{xcolor}
\usepackage{braket}
\usepackage{hyperref}
\usepackage[normalem]{ulem}

\begin{document}
\title{Photon-induced droplet-like bound states in one-dimensional qubit array}
\author{J. Talukdar}
\address{Homer L. Dodge Department of Physics and Astronomy,
  The University of Oklahoma,
  440 W. Brooks Street,
  Norman,
Oklahoma 73019, USA}
\address{Center for Quantum Research and Technology,
  The University of Oklahoma,
  440 W. Brooks Street,
  Norman,
Oklahoma 73019, USA}
\author{D. Blume}
\address{Homer L. Dodge Department of Physics and Astronomy,
  The University of Oklahoma,
  440 W. Brooks Street,
  Norman,
Oklahoma 73019, USA}
\address{Center for Quantum Research and Technology,
  The University of Oklahoma,
  440 W. Brooks Street,
  Norman,
Oklahoma 73019, USA}
\date{\today}

\begin{abstract}
We consider an array of $N_e$ non-interacting qubits or emitters that are coupled to a one-dimensional cavity array with tunneling energy $J$ and non-linearity of strength $U$. The number of cavities is assumed to be larger than the number of qubits. Working in the two-excitation manifold, we focus on the bandgap regime where the energy of two excited qubits is off-resonant 
with the two-photon bound state band. A two-step adiabatic elimination of the photonic degrees of freedom gives rise to a one-dimensional spin Hamiltonian with effective interactions; specifically, the Hamiltonian features constrained single-qubit hopping and pair hopping interactions not only between nearest neighbors but also between next-to-nearest and next-to-next-to-nearest spins. For a regularly arranged qubit array, we identify parameter combinations for which the system supports novel droplet-like bound states whose characteristics depend critically on the pair hopping. The droplet-like states can be probed dynamically. The bound states identified in our work for off-resonance conditions are distinct from localized hybridized states that emerge for on-resonance conditions.
\end{abstract}
\maketitle

\section{Introduction}
\label{sec_introduction}

Qubits or, more generally, few-level emitters coupled to a cavity array provide a platform with which to investigate fundamental aspects of matter-light interactions. Topics of interest include the generation of photon-mediated entanglement between non-interacting separated qubits~\cite{ref_sinha,ref_chiral-zoller,ref_Haakh,ref_chaos}, of ultrastrong matter-light interactions~\cite{ref_henriet,ref_wolf,ref_hwang,ref_altintas,ref_mahmoodian,ref_kockum,ref_diaz}, of broad matter-light hybrid bound states~\cite{ref_ripoll-1,ref_ripoll-2,ref_shi,ref_rabl_atom-field,ref_pcw,ref_cirac_many-body,ref_jugal_paper-0}, and of effective photon-photon interactions~\cite{ref_google-group, ref_gorshkov_ph-ph,ref_jeanic_ph-ph, ref_pohl_ph-ph}. Photonic baths have been realized using nanophotonic wave guides~\cite{ref_roy,ref_nano-shah,ref_palma,ref_wg_many-bound,ref_few-ph,ref_reitz,ref_yalla,ref_hood}, superconducting resonators~\cite{ref_resonator,ref_sqbit,ref_photon-mat_cqed, ref_cqed-jon-simon}, and plasmonic waveguides~\cite{ref_plasmom}. Qubit realizations include Rydberg atoms~\cite{ref_saffman, ref_levine}, quantum dots~\cite{ref_qdot}, and transmon qubits~\cite{ref_astafiev,ref_hoi,ref_mlynek,ref_mirhosseini,ref_sundaresan}. 

It was recently shown that the addition of a Kerr-like non-linearity to the tight-binding Hamiltonian, which accounts for the tunnel-coupling of the single-mode cavities, leads to intriguing and qualitatively novel phenomena if the energy of two excited qubits is tuned to be in resonance with the two-photon bound state band that exists due to the Kerr-like non-linearity~\cite{ref_rabl_non-linear, ref_jugal_paper-1, ref_jugal_paper-2}. For two qubits initialized in their excited state, e.g., the non-trivial mode structure of the bath, i.e., the cavity array with non-linearity, was shown to support emission dynamics that ranges from exponential decay to fractional populations to Rabi oscillations~\cite{ref_jugal_paper-1, ref_jugal_paper-2}. For many qubits, supercorrelated radiance was predicted~\cite{ref_rabl_non-linear}. This work instead investigates the off-resonant 
or band-gap
regime~\cite{ref_john}
within the framework of Schr\"{o}dinger quantum mechanics. To reduce the high-dimensional Hilbert space to a physically intuitive and numerically more tractable model, effective constrained single-qubit and pair hopping interactions are derived through a two-step procedure that adiabatically eliminates single- and two-photon processes. The resulting effective one-dimensional spin Hamiltonian, which lives in the two-excitation manifold (i.e., two flipped spins), is shown to capture the key features of the full Hamiltonian.

The effective constrained single- and two-qubit hopping interactions, which are derived under the assumption that the coupling strength $g$ between an emitter and a cavity is small compared to the tunneling energy $J$, are directly proportional to $g^2$ and $g^4$, respectively. Even though the scaling of the effective interactions with $g$ suggests that the single-qubit hopping dominates over the two-qubit hopping, we identify a parameter regime where the latter, which depends on the non-linearity $U$, impacts the eigenstate characteristics appreciably. Specifically, the pair hopping interaction favors localization of excited qubits in or near the middle of the 
qubit array, giving rise to a new class of droplet-like bound states. These bound states are distinct from two-string bound states that exist, e.g., in the XXX spin Hamiltonian that is solvable via the Bethe ansatz~\cite{ref_XXX-bethe, ref_wang-nature}. Unlike Hamiltonian that are tractable via the Bethe ansatz, our emergent one-dimensional spin model features non-negligible nearest-neighbor, next-to-nearest-neighbor, and next-to-next-to-nearest-neighbor interactions. It is shown that the radiation dynamics, if initiated from an initial state that contains two qubit excitations but no photons, depends strongly on how the two qubit excitations are distributed among all possible two-qubit excitation eigenkets. A fully symmetric initial state is shown to induce oscillatory dynamics between the 
droplet-like
ground 
state and a scattering state. 
Dependence of the dynamics on the initial state is, of course, a well known phenomenon that has, e.g., been exploited in the study of phase transitions and critical points as well as in  sensing applications.

The remainder of this article is organized as follows. Section~\ref{sec_theoretical-background} introduces the system Hamiltonian and the reduction of the Hilbert space to the qubit degrees of freedom. Section~\ref{sec_stationary-solutions} shows that the effective qubit Hamiltonian supports a new class of liquid-like 
or droplet-like bound states. Section~\ref{sec_dynamics} illustrates that these droplet-like states can be probed dynamically. Last, a summary and outlook are provided in Sec.~\ref{sec_conclusion}. 

\section{Derivation of Effective Qubit Hamiltonian}
\label{sec_theoretical-background}
Section~\ref{sec_hamiltonian} introduces the total Hamiltonian $\hat{H}$ of the matter-light hybrid system. Focusing on the band gap regime of the photonic lattice, Sec.~\ref{sec_approx_spin} derives the effective spin Hamiltonian $\hat{H}_{\text{spin}}$. 

\subsection{Total Hamiltonian $\hat{H}$}
\label{sec_hamiltonian}
\begin{figure}[t]
\includegraphics[width=0.48\textwidth]{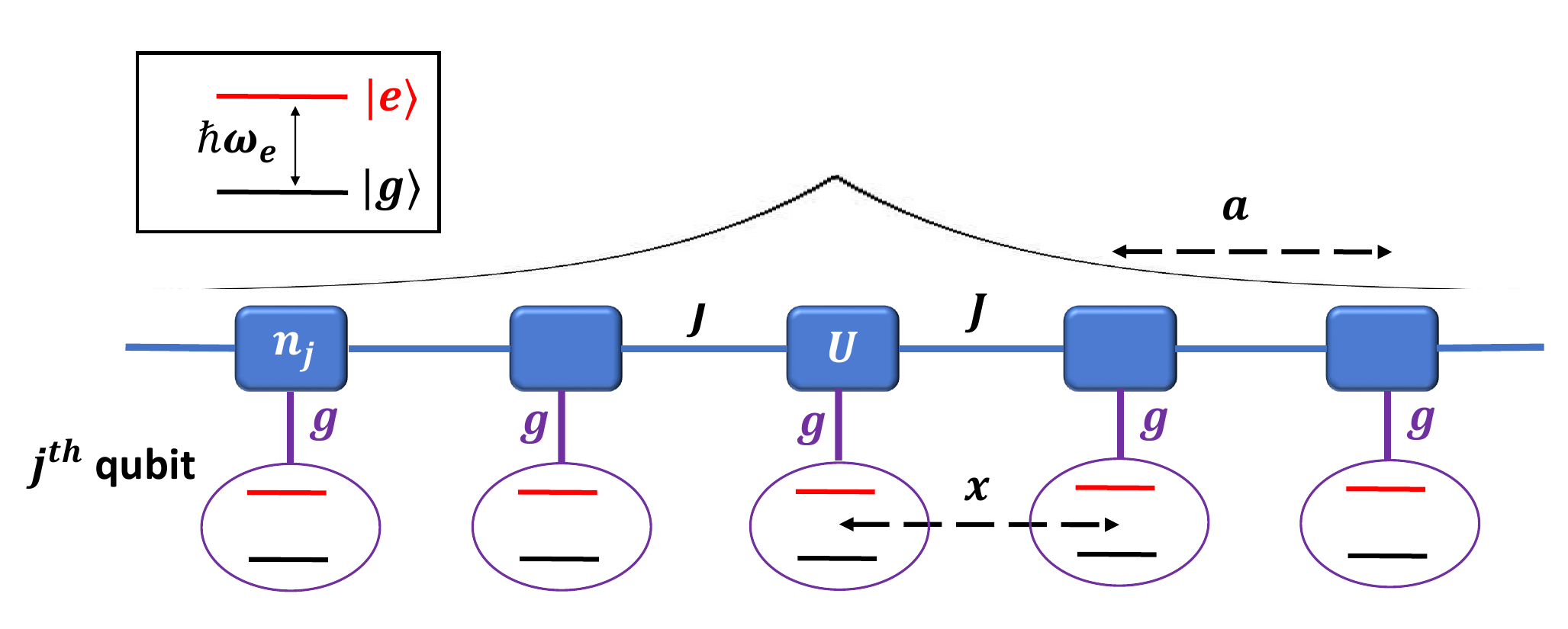}
\caption{Schematic of the set up. The $j$th qubit is coupled with strength $g$ to the $n_j$th cavity of a one-dimensional cavity array (blue boxes) with lattice spacing $a$. The cavities are tunnel-coupled to nearest neighbors with strength $J$ (blue lines between two neighboring blue boxes). For more than one photonic excitation, there exists an onsite interaction $U$ between photons. As a result of the onsite interaction, the cavity array (bath) supports two-photon bound states. One of these is shown by the black line above the cavity array. The distance between two neighboring qubits is denoted by $x$. In the schematic, $x$ is equal to $a$; values of $x/a=0$ and $2$ are also discussed in this work. The top-left rectangular box illustrates a qubit, i.e., a two-level system with a transition energy $\hbar\omega_e$ between the ground state $\ket{g}$ and the excited state $\ket{e}$.}
\label{fig0}
\end{figure}

The total Hamiltonian $\hat{H}$ reads
\begin{eqnarray}
\label{eq_ham_total}
\hat{H} = \hat{H}_{\text{qubit}} + \hat{H}_{\text{bath}} + \hat{H}_{\text{qubit-bath}},
\end{eqnarray}
where $\hat{H}_{\text{qubit}}$ is the Hamiltonian of the uncoupled qubits, $\hat{H}_{\text{bath}}$ the bath Hamiltonian, and $\hat{H}_{\text{qubit-bath}}$ the qubit-bath coupling Hamiltonian. The qubit system consists of $N_e$ qubits with a transition energy of $\hbar\omega_e$ between the ground state $\ket{g}_j$ and the excited state $\ket{e}_j$ of the $j$th qubit (see purple ovals and rectangular box in top-left corner in Fig.~\ref{fig0}). We are interested in the regime where the qubits form a regularly arranged finite lattice ($N_e$ finite and much greater than $1$). The qubit Hamiltonian $\hat{H}_\text{qubit}$ is given by
\begin{eqnarray}
\label{eq_ham_sys}
\hat{H}_{\text{qubit}}=
\frac{\hbar \omega_e}{2}
\sum_{j=1}^{N_e} 
(\hat{\sigma}_{j}^z + \hat{I}_j),
\end{eqnarray}
where $\hat{\sigma}_{j}^z=\ket{e}_j\bra{e}-\ket{g}_j\bra{g}$ and $\hat{I}_{j}^z=\ket{e}_j\bra{e}+\ket{g}_j\bra{g}$. 

\begin{figure}[t]
\includegraphics[width=0.48\textwidth]{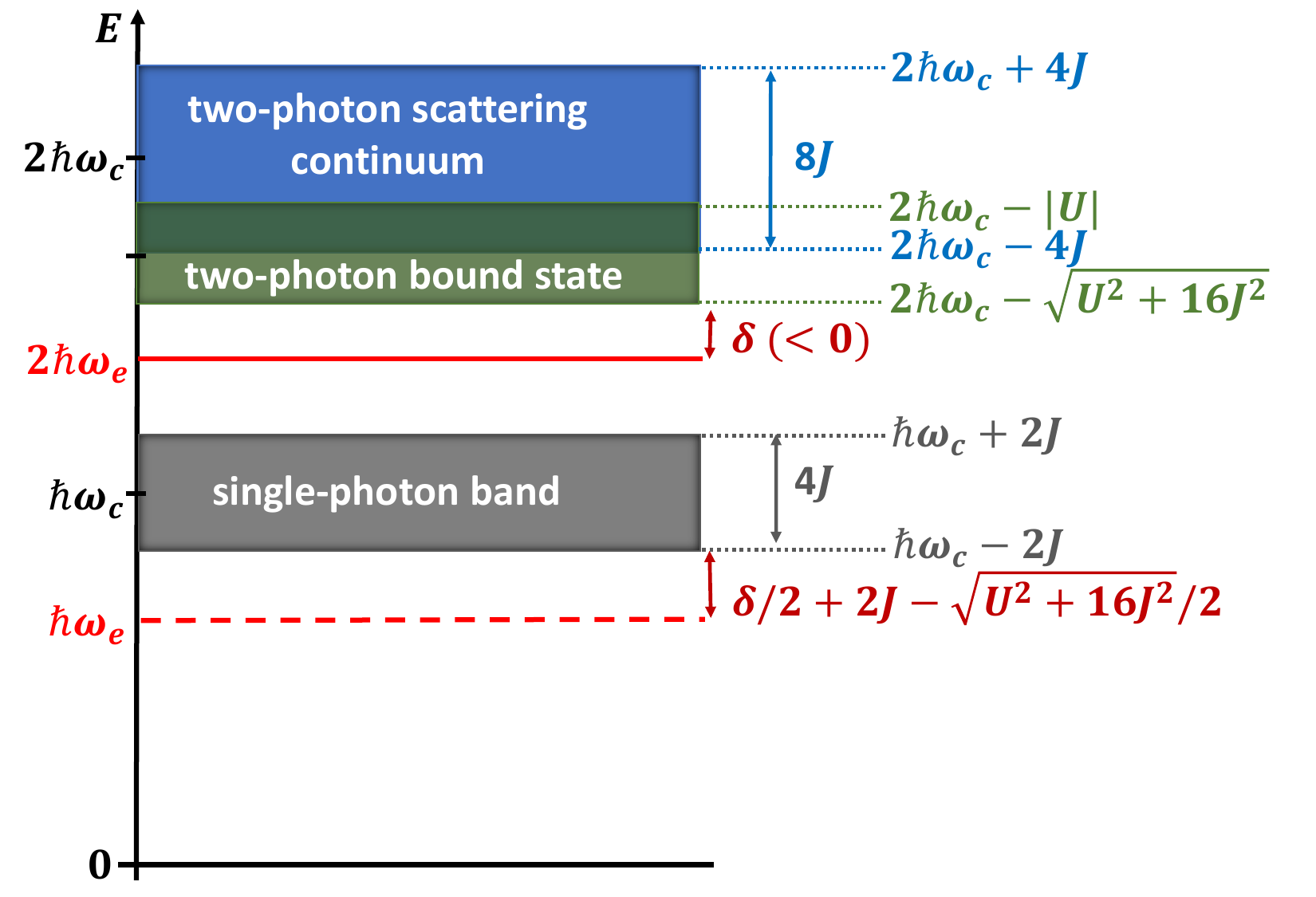}
\caption{Schematic of the energy bands for the one- and two-excitation manifolds for fixed $\omega_c$, $\omega_e$, and $U$. This work focuses on the band gap regime, i.e., negative detunings $\delta$. Adiabatic elimination of the gray single-photon and green two-photon bound state bands introduces the effective interactions $W$ and $Y$ (see Fig.~\ref{fig_W-Y_schematic} for an illustration of these interactions), respectively, between qubit groups. The two-photon scattering continuum is far off-resonant and does not play a role. Explicit expressions for the energy bands can 
be
found, e.g., in Ref.~\cite{ref_valiente}.}
\label{fig_band}
\end{figure}

The bath Hamiltonian $\hat{H}_{\text{bath}}$ is a one-dimensional tight-binding Hamiltonian with non-linearity $U$,
\begin{eqnarray}
\label{eq_bath_ham}
\hat{H}_{\text{bath}} =&&
\hbar \omega_c  \sum_{n=1}^{N} 
 \hat{a}_n^{\dagger} \hat{a}_n 
-J  \sum_{n=1}^{N} \left(
\hat{a}_n^{\dagger} \hat{a}_{n+1} + \hat{a}_{n+1}^{\dagger} \hat{a}_{n} 
\right) \nonumber \\
&&+\frac{U}{2} 
\sum_{n=1}^{N}
 \hat{a}_n^{\dagger}\hat{a}_n^{\dagger}\hat{a}_n\hat{a}_n,
\end{eqnarray}
where $\hat{a}^{\dagger}_n$ and $\hat{a}_n$, respectively, create and destroy a photon at the $n$th cavity (blue box in Fig.~\ref{fig0}). In our calculations, the number of cavities $N$ is chosen such that the results are independent of $N$; we find that $N=501$ is sufficiently large for the $N_e$ considered. In Eq.~(\ref{eq_bath_ham}), $\hbar\omega_c$ is the single-mode photon energy, $J$ ($J>0$) denotes the tunneling energy of the tunnel coupled cavities, and $U$ is the non-linear onsite interaction. The Kerr-like non-linearity in Eq.~(\ref{eq_bath_ham}) corresponds to effectively repulsively interacting photon pairs ($U > 0$) or effectively attractively interacting photon pairs ($U < 0$). In our work, we consider a negative $U$, which gives rise to two-photon bound states $\psi_{K,b}$ with center-of-mass wave vector $K$ and energy $E_{K,b}$, in addition to the two-photon scattering continuum (blue and dark green regions in Fig.~\ref{fig_band})~\cite{ref_molmer1, ref_molmer2, ref_valiente, ref_petrosyan}. The black line in Fig.~\ref{fig0} shows a sketch of a two-photon bound state wave function $\psi_{K,b}$ that extends over several lattice sites. Accounting for all allowed center-of-mass wave vectors $K$, the two-photon bound states give rise to an energy band (green and dark green regions in Fig.~\ref{fig_band}). For large values of the onsite interaction strength $|U|$ ($|U|/J>4$), the two-photon bound state band does not overlap with the two-photon scattering continuum. For $|U|/J=1$, as considered in this paper, the upper part of the two-photon bound state band overlaps with the lower part of the two-photon scattering continuum (the overlap region is shown in dark green in Fig.~\ref{fig_band}). The difference between the energy $2\hbar\omega_e$ of the two-qubit excited state and the $K=0$ two-photon bound state energy $E_{0,b}$, which coincides with the bottom of the two-photon bound state band, defines the detuning $\delta$,
\begin{equation}
    \delta=2\hbar\omega_e-E_{0,b}.
\end{equation}
The band gap regime, which is the focus of the present work, is characterized by negative detunings $\delta$. 

The qubits are coupled to the photons through the system-bath or qubit-bath Hamiltonian $\hat{H}_{\text{qubit-bath}}$,
\begin{eqnarray}\label{eq_ham_sb}
\hat{H}_{\text{qubit-bath}}= 
g \sum_{j=1}^{N_e}
\left(
\hat{a}_{n_j} \hat{\sigma}_j^+ + \hat{a}_{n_j}^{\dagger}  \hat{\sigma}_j^-
\right),
\end{eqnarray}
where $\hat{\sigma}_j^+$ is the raising operator ($\hat{\sigma}_j^+=\ket{e}_j\bra{g}$) and $\hat{\sigma}_j^-$ the lowering operator ($\hat{\sigma}_j^-=\ket{g}_j\bra{e}$) of the $j$th qubit. The label $n_j$ can take any value between 1 and $N$. In this work, the qubits are assumed to be arranged in a regular pattern with spacing $x$, where $x/a=n_j-n_{j-1}$. 
Related works considered regularly placed impurity qubits coupled to an atomic array~\cite{ref_atw-1,ref_atw-2}.
Our figures concentrate on $x/a=1$. For reference, a larger qubit spacing $x/a=2$ as well as the case where the qubits are all coupled to the same cavity ($x/a=0$) are discussed in the text. Since the counter rotating terms are excluded in Eq.~(\ref{eq_ham_sb}), our treatment is restricted to the weak coupling regime, i.e., $g \ll J$. The requirement that single- and two-photon processes are off-resonant [$|(\hbar\omega_c-2J)-\hbar\omega_e|>g$ and $|\delta| > g$] can, for negative $\delta$ as considered in this work, be combined 
into
one equation, namely
\begin{equation}
\label{eq_g-U_single-ae}
|U|>4J\sqrt{\left(1+\frac{g}{4J}\right)^2-1}.  
\end{equation}
For fixed $U/J$, Eq.~(\ref{eq_g-U_single-ae}) puts an upper limit on $g/J$. Conversely, for fixed $g/J$, Eq.~(\ref{eq_g-U_single-ae}) puts a lower limit on $|U|/J$.

The total Hamiltonian conserves the number of total excitations (sum of qubit and photonic excitations)~\cite{ref_ripoll-1,ref_ripoll-2, ref_rabl_non-linear,ref_shi,ref_rabl_atom-field}. As a consequence, the Hilbert spaces with $0$, $1$, $2$, $\dots$ total excitations are decoupled. This work focuses on the two-excitation manifold.

\subsection{Effective spin Hamiltonian $\hat{H}_{\text{spin}}$}
\label{sec_approx_spin}
As mentioned above, we focus on negative detunings such that the energy of two excited qubits is in resonance with the band gap. We find that the band gap physics in the two-excitation manifold is well described by the spin Hamiltonian $\hat{H}_{\text{spin}}$, which is derived by adiabatically eliminating the photon degrees of freedom in a two-step process (see Appendix~\ref{append_A} for details). We emphasize that the approach taken here is distinct from the master equation approach pursued in Ref.~\cite{ref_rabl_non-linear}. The first step is, in spirit, identical to prior work~\cite{ref_rabl_non-linear,ref_jugal_paper-1, ref_jugal_paper-2}. Neglecting the two-photon scattering continuum and adiabatically eliminating the single-photon states, effective constrained single-qubit hopping interactions of strength $W_{jl}$ (see $\hat{H}_{\text{single}}$ below), effective interactions between states with two and no qubit excitations [$F_{K,b}$ in Eq.~(\ref{eq_fsubcapk})], and effective interactions between two two-photon bound states with wave vector $K$ and $K'$ [$G_{K,K'}$ in Eq.~(\ref{eq_GKKprime})] arise. While the latter two interactions were discussed in Refs.~\cite{ref_rabl_non-linear,ref_jugal_paper-1, ref_jugal_paper-2}, the effective qubit hopping interaction was not. The reason is that Refs.~\cite{ref_rabl_non-linear,ref_jugal_paper-1, ref_jugal_paper-2} focused on $N_e=2$ ($\hat{H}_{\text{single}}$ vanishes for $N_e=2$). The hopping Hamiltonian $\hat{H}_{\text{single}}$ reads
\begin{eqnarray}
\label{eq_H-hopping}
\nonumber\hat{H}_{\text{single}}=\frac{1}{2}\sum_{i,j,l=1}^{N_e}\Big(W_{jl}\hat{\sigma}^+_i\hat{\sigma}^+_j\hat{\sigma}^-_i\hat{\sigma}^-_l+ \\W_{il}\hat{\sigma}^+_i\hat{\sigma}^+_j\hat{\sigma}^-_l\hat{\sigma}^-_j \Big).
\end{eqnarray}
Since the triple sum includes terms where two or three of the indices are equal, the order of the operators in Eq.~(\ref{eq_H-hopping}) is important. As discussed in more detail below, $\hat{H}_{\text{single}}$ describes constrained single-qubit hopping
or constrained flip-flop interactions. We find that the effective interactions $G_{K,K'}$ contribute negligibly to the band gap physics considered in this 
work; thus, they are set to zero.

Calculations that treat the full Hamiltonian $\hat{H}$ show that the photonic contribution to the eigenstates is smaller than $10\%$ for the parameter combinations considered in this work. This motivates  
our
second approximation, namely, the adiabatic elimination of the states $\hat{B}_K^{\dagger}\ket{g,\cdots,g, \text{vac}}$, i.e., basis kets that describe a photon pair with wave vector $K$, with the qubits in the ground state. Step two yields the effective spin Hamiltonian $\hat{H}_{\text{spin}}$ (see Appendix~\ref{append_A} for details),
\begin{eqnarray}
\label{eq_H_eff2}
 \hat{H}_{\text{spin}}= \hat{H}_{\text{single}}+\hat{H}_{\text{pair}},
\end{eqnarray}
where
\begin{eqnarray}
\label{eq_H-Y}
\hat{H}_{\text{pair}}=\sum_{i=1}^{N_e-1}\sum_{ j=i+1}^{N_e}\sum_{l=1}^{N_e-1}\sum_{h=l+1}^{N_e}Y_{ij,lh}\hat{\sigma}^+_i\hat{\sigma}^+_j\hat{\sigma}^-_l\hat{\sigma}^-_h.
\end{eqnarray}
The effective four-qubit (or two-qubit hopping) interactions $Y_{ij,lh}$ emerge from the interactions $F_{K,b}$ (see below). As might be expected naively, $W_{jl}$ and $Y_{ij,lh}$ are directly proportional to $g^2$ and $g^4$, respectively, since they emerge as a consequence of the first and second adiabatic elimination steps, respectively. The effective spin Hamiltonian $\hat{H}_{\text{spin}}$ is independent of the photonic degrees of freedom. The characteristics of the cavity array and the geometric arrangement of the qubits (i.e., the value of $x$) enter through the interaction strengths $W_{jl}$ and $Y_{ij,lh}$. 

\begin{figure}
\includegraphics[width=0.51\textwidth]{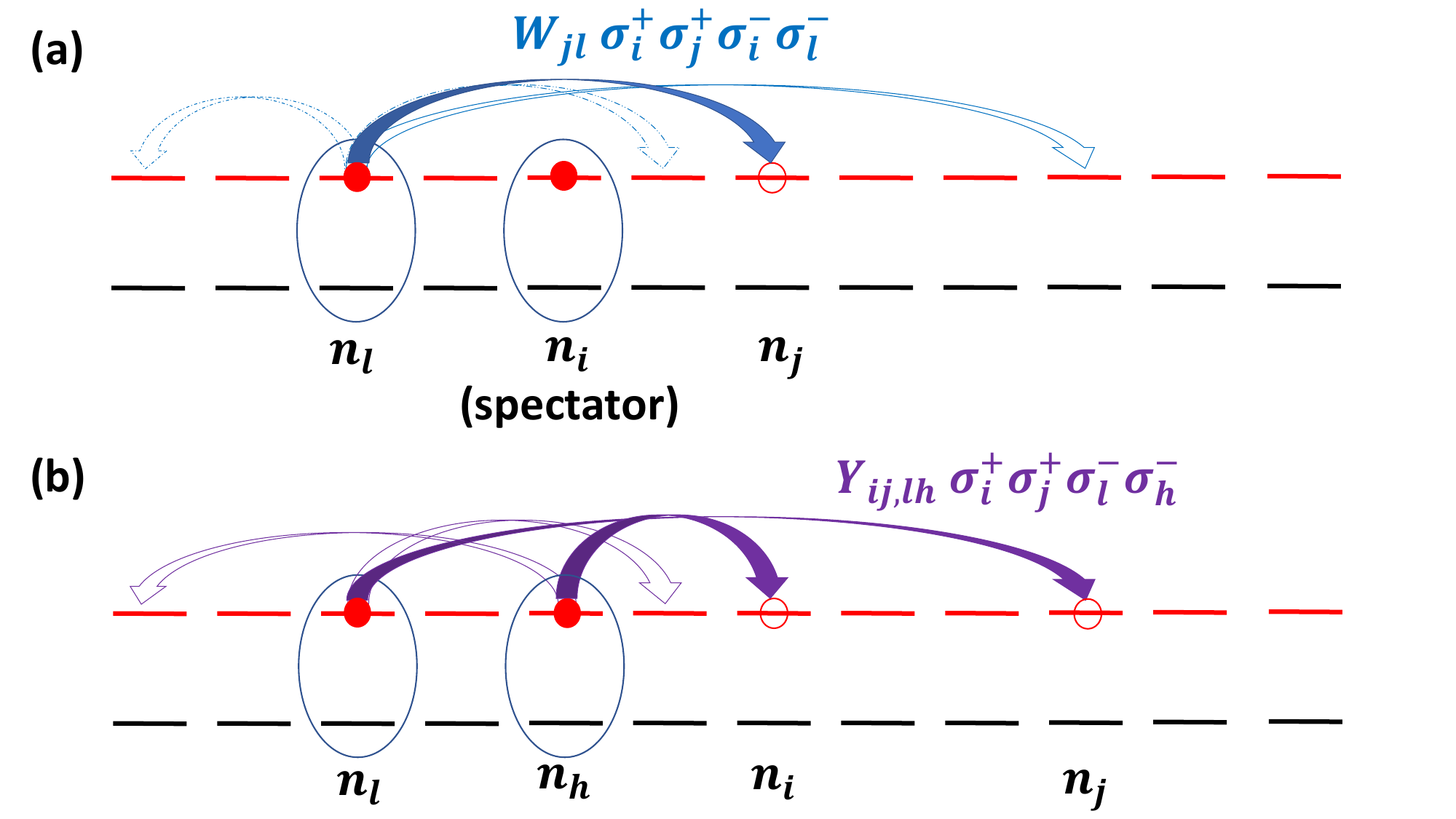}
\caption{Schematic of constrained single-qubit hopping interaction $W$ and pair hopping interaction $Y$ entering into $\hat{H}_{\text{spin}}$. (a) The term $W_{jl}\hat{\sigma}^+_i\hat{\sigma}^+_j\hat{\sigma}^-_i\hat{\sigma}^-_l$ (here illustrated assuming $i\neq j\neq l$) describes the annihilation of an excitation at the $l$th qubit and the creation of an excitation at the $j$th qubit (solid blue arrow). This corresponds to the hopping of an excitation with strength $W_{jl}$, with the excitation at the $i$th qubit acting as a ``spectator", i.e., the single-qubit hopping is only allowed if qubit $i$ is excited. (b) The term $Y_{ij,lh}\hat{\sigma}^+_i\hat{\sigma}^+_j\hat{\sigma}^-_l\hat{\sigma}^-_h$ (here illustrated assuming $i\neq j\neq l\neq h$) describes the annihilation of excitations at qubits $l$ and $h$, and the creation of excitations at qubits $i$ and $j$ (solid purple arrows). This corresponds to the hopping of a pair of excitations with strength $Y_{ij,lh}$. The open blue and open purple arrows show selected additional constrained hopping and pair hopping interactions, respectively.}
\label{fig_W-Y_schematic}
\end{figure}

We now highlight selected properties of the single- and two-qubit hopping interactions. Figure~\ref{fig_W-Y_schematic}(a) illustrates the constrained single-qubit hopping interaction $W_{jl}\hat{\sigma}^+_i\hat{\sigma}^+_j\hat{\sigma}^-_i\hat{\sigma}^-_l$. The term ``constrained'' is used since the hopping of the excitation from qubit $l$ to qubit $j$ ($\hat{\sigma}^+_j\hat{\sigma}^-_l$ piece) depends on the number of excitations at qubit $i$ ($\hat{\sigma}^+_i\hat{\sigma}^-_i$ piece; in this example, we assume $i\neq j$ and $i\neq l$). If qubit $i$ is excited, hopping from qubit $l$ to qubit $j$ occurs with strength $W_{jl}$. If, in contrast, qubit $i$ is not excited, hopping from qubit $l$ to qubit $j$ does not take place. We refer to the excited qubit $i$ as a spectator. 
We emphasize that
our treatment does not assume that the system is in the Markovian regime. 
After the first adiabatic elimination, basis kets with two excited qubits are coupled to each other via $\hat{H}_{\text{single}}$ if they contain a common excited qubit. The second adiabatic elimination leaves $\hat{H}_{\text{single}}$ unchanged. Thus, in the Hilbert space spanned by the $N_e(N_e-1)/2$ two-excitation qubit states, $\hat{H}_{\text{single}}$ couples each basis ket that contains two excited qubits to $2(N_e-2)$ other basis kets as well as to itself. While we 
refer
to $W(0)$ as onsite hopping interaction, it is also known as ``self interaction'' 
or ``self energy'' 
(see, e.g., Ref.~\cite{ref_rabl_non-linear}).

\begin{figure}
\includegraphics[width=0.47\textwidth]{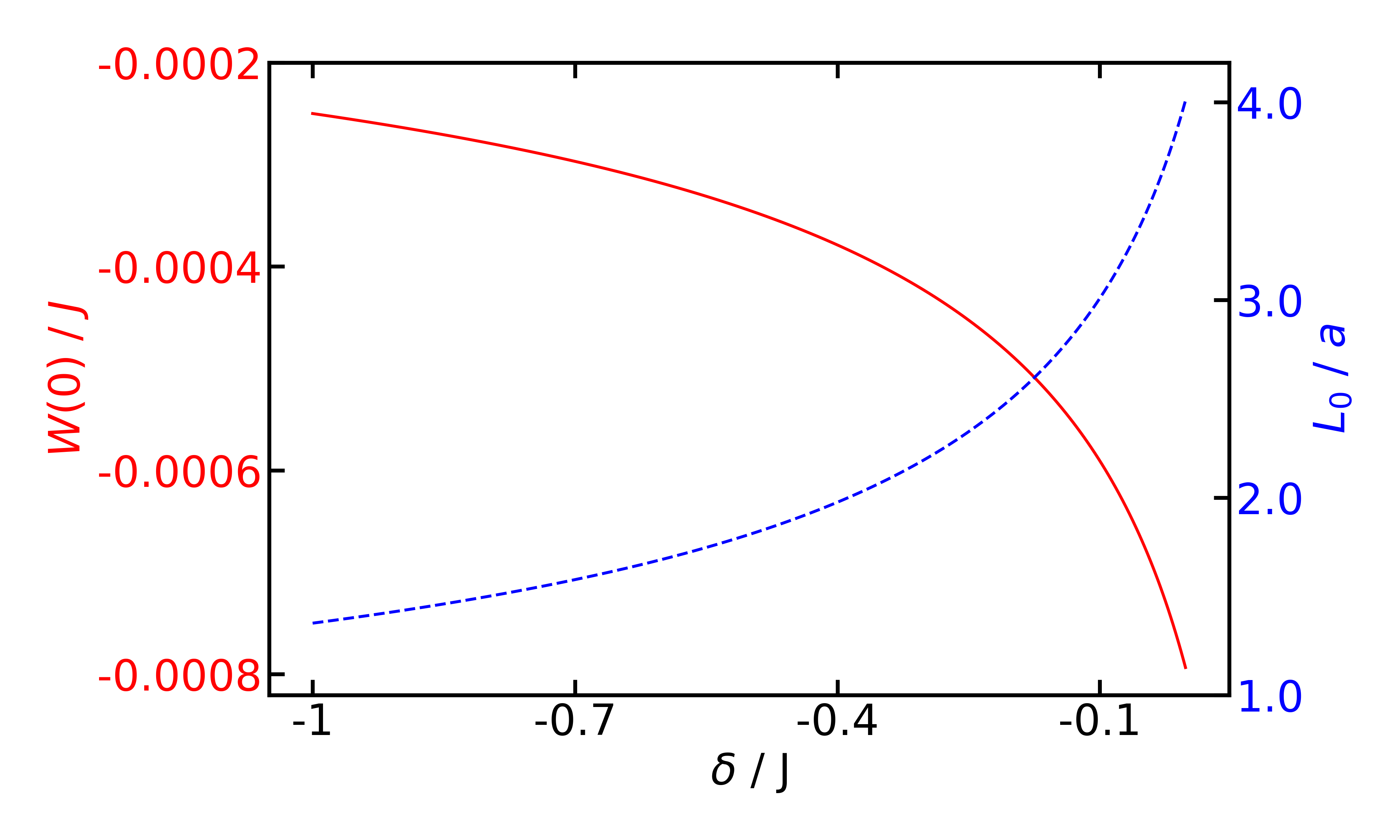}
\caption{The red solid and blue dashed lines show the dimensionless on-site hopping energy $W(0)/J$ (left axis) and length $L_0/a$ (right axis) as a function of the detuning $\delta/J$ for $g/J=1/50$ and $U/J=-1$.
}
\label{fig_w-scale}
\end{figure}

The strength $W_{jl}$,
\begin{eqnarray}
\label{eq_W}
W_{jl}=W(0) \exp(-|n_j-n_l|a/L_0),
\end{eqnarray}
of the constrained single-qubit hopping interaction falls off exponentially as a function of $|n_j-n_l|a$, i.e., the difference between the cavities $n_j$ and $n_l$ that the qubits $j$ and $l$ are coupled to. The onsite hopping energy $W(0)$ and length $L_0$ read
\begin{equation}
    W(0)=-\frac{2J\left(\frac{g}{2J}\right)^2}{\sqrt{\left(\frac{\Delta}{2J}\right)^2-1}}
\end{equation} 
and
\begin{equation}
    L_0=-\frac{a}{\ln\left(\frac{\Delta}{2J}-\sqrt{\left(\frac{\Delta}{2J}\right)^2-1}\right)},
\end{equation}
respectively, where
\begin{equation}
    \Delta=\hbar\left(\omega_c-\omega_e\right)=\frac{1}{2}\left(-\delta+4J\sqrt{1+\left(\frac{U}{16J}\right)^2}\right).
\end{equation} 
Figure~\ref{fig_w-scale} shows the onsite hopping energy $W(0)$ and length $L_0$ for fixed $g/J$ and $U/J$ as a function of the dimensionless detuning $\delta/J$. It can be seen that $W(0)/J$ is negative and that the magnitude of $W(0)$ increases with decreasing $|\delta/J|$. Larger $|W(0)|$ (note, Fig.~\ref{fig_w-scale} shows $W(0)$ as opposed to $|W(0)|$) are accompanied by larger $L_0$. For the detuning considered in this work ($|\delta/J|\ll 1$), $W_{jl}$ is---for $x/a=1$---appreciable not only for nearest neighbor hopping but also for next-to-nearest and next-to-next-to-nearest neighbor hopping.

\begin{figure}[h]
\begin{center}
\includegraphics[width=0.50\textwidth]{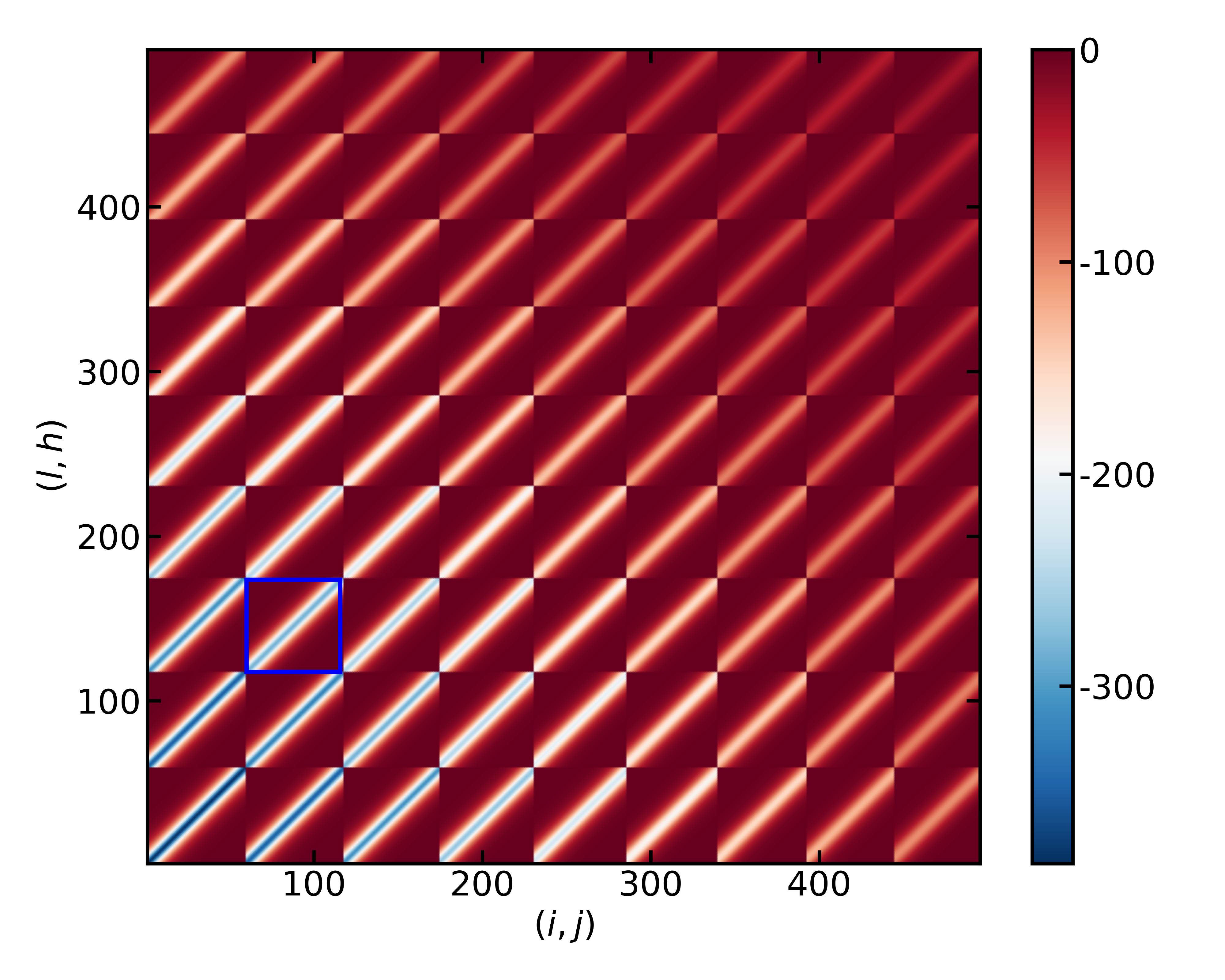}
\end{center}
\caption{Contour plot of the effective dimensionless interaction $Y_{ij,lh}J^3/g^4$ for $U/J=-1$, $\delta/J=-1/50$, $N_e=60$, and $x/a=1$. The $x$- and $y$-axis are labeled by the index pairs ($i, j$) and ($l, h$); the plot includes all ($i, j$) and ($l, h$) pairs with $|j-i|\leq 9$ and $|h-l|\leq 9$. In each block, the separation (i.e., $j-i$ and $h-l$) between the two qubit excitations is fixed while the ``center-of-mass coordinates'' [i.e., $(i+j)/2$ and $(l+h)/2$] are changing. As an example, the blue rectangle corresponds to a block with $|j-i|=2$ and $|h-l|=3$. Values of 
($x$, $y$) = ($100$, $150$), e.g., 
correspond
to $(i, j)= (41, 43)$ and $(l, h)= (33, 36)$.}
\label{fig_Y}
\end{figure}

Next, we discuss the effective pair hopping interaction $Y_{ij,lh}$. Figure~\ref{fig_W-Y_schematic}(b) illustrates $Y_{ij,lh}\hat{\sigma}^+_i\hat{\sigma}^+_j\hat{\sigma}^-_l\hat{\sigma}^-_h$, which annihilates excitations at the $l$th and $h$th qubit and creates excitations at the $j$th and $i$th qubit. The effective pair hopping interaction $Y_{ij,lh}$ is given by 
\begin{eqnarray}
Y_{ij,lh}=-\frac{g^4}{NJ^2}\sum_K\frac{F_{K,b}(n_i,n_j)F_{K,b}^*(n_l,n_h)}{\Delta_{K,b}},
\end{eqnarray}
where $F_{K,b}$ is given in Eq.~(\ref{eq_fsubcapk}). As $W_{jl}$, $Y_{ij,lh}$ is negative. The pair hopping interaction $\hat{H}_\text{pair}$ couples each excited qubit pair to all other excited qubit pairs. Figure~\ref{fig_Y} shows the interaction $Y_{ij,lh}$ for $x/a=1$ as functions of the pairs ($i,j$) and ($l,h$) for $N_e=60$. Specifically, the indices that specify the states $\hat{\sigma}_i^+\hat{\sigma}_j^+\ket{g,\cdots,g}$ are organized based on the separation between the excited qubits, i.e., $|j-i|$. In a qubit array with $N_e$ qubits, there are $N_e-1$ basis states with a separation of $|j-i|=1$, $N_e-2$ basis states with a separation of $|j-i|=2$, and so on. The ``lower left block" corresponds to $|j-i|=|h-l|=1$ [the pairs ($i,j$) and ($l,h$) both take the values ($1,2$), ($2,3$),$\cdots$, ($59,60$)]. The ``upper right block'' corresponds to $|j-i|=|h-l|=9$ [the pairs ($i,j$) and ($l,h$) both take the values ($1,10$), ($2,11$),$\cdots$, ($51,60$)]. Note that Fig.~\ref{fig_Y} only considers a subset of pairs, i.e., $|j-i|\leq 9$ and $|h-l|\leq 9$. Within each block, the interaction is most negative along the diagonal and falls off approximately Lorentzian as one moves away from the diagonal. Moreover, starting with the block in the lower left corner, the interactions on the diagonal are less negative as one moves to blocks characterized by larger separations. 

A key characteristic of the interaction $Y_{ij,lh}$ is that it is---within each block---constant along the diagonal, along the off-diagonal, and so on. The fall-off of the interactions as one moves away from the diagonal within each block indicates that the $Y_{ij,lh}$ interaction depends on the actual locations of the involved qubits in the spin chain. This implies that it is energetically more favorable for two excitations to be located in the middle of the chain than at the edge of the chain since the pair can hop to the left and to the right when located at the center and only to one side when located at the edge. This location dependence is critical for the formation of the droplet-like states discussed in the next section.  

\section{Stationary Solutions}
\label{sec_stationary-solutions}

Since we are working in the regime where $g/J\ll 1$, it might be expected naively that the constrained single-qubit hopping term $\hat{H}_{\text{single}}$, which is directly proportional to $g^2$, dominates over the pair hopping term $\hat{H}_{\text{pair}}$, which is directly proportional to $g^4$. While this is, indeed, the case in an appreciable portion of the parameter space, we show that there exists a parameter window in which the pair hopping interaction qualitatively changes the system characteristics.
It is noted that a fourth-order two-photon virtual process, which is proportional to $g^4$, was observed experimentally in transmon qubits coupled to a photonic crystal~\cite{ref_sundaresan}.
Specifically, this section shows that the $Y$-term has a ``pinning effect'' that leads to the emergence of liquid-
or droplet-like 
bound states. 
Droplet states are self-bound and incompressible, and their excitation spectrum can be divided into compressional and surface modes~\cite{casas1990}. We will show that the states referred to as droplet-like in this work  are incompressible (their size is not solely set by the extent of the emitter array but by the entirety of system parameters). Moreover, the ground state is accompanied by a sequence of excitations that resemble compressional modes.
While our analysis is based on the approximate spin Hamiltonian $\hat{H}_{\text{spin}}$, we checked that this Hamiltonian captures the key features of the full system Hamiltonian $\hat{H}$ qualitatively and in many cases even quantitatively correctly. The main advantage of using $\hat{H}_{\text{spin}}$ comes from the fact that it allows for a transparent interpretation of the results, in addition to being an interesting model in its own right.

\begin{figure}[h]
\begin{center}
\includegraphics[width=0.47\textwidth]{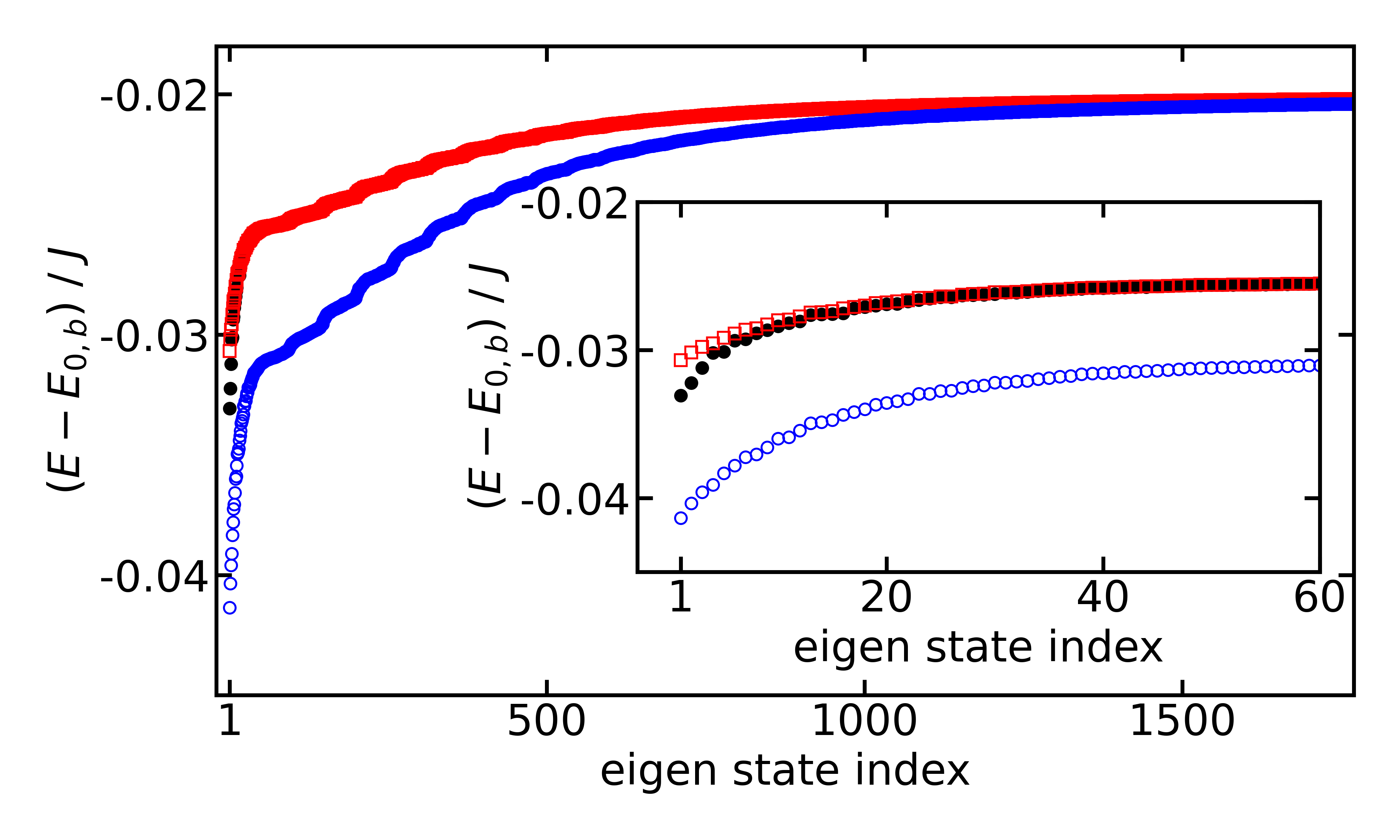}
\end{center}
\caption{Eigenenergy, measured with respect to $E_{0,b}$, for $N_e=60$, $g/J=1/50$, $U/J=-1$, $\delta/J=-1/50$, and $x/a=1$ as a function of the state index. The black filled circles, red open squares, and blue open circles show the energy for $\hat{H}_{\text{spin}}$, $\hat{H}_{\text{single}}$, and $\hat{\Tilde{H}}_{\text{single}}$, respectively. Inset: Blow-up of the lower part of the energy spectrum.}
\label{fig_energy}
\end{figure}

We start by setting $\hat{H}_{\text{pair}}=0$. We find it useful to compare $\hat{H}_{\text{single}}$ to the unconstrained one-qubit hopping Hamiltonian $\hat{\Tilde{H}}_{\text{single}}$, where $\hat{\Tilde{H}}_{\text{single}}=2\sum_{i,j=1}^{N_e}W_{ij}\hat{\sigma}_i^+\hat{\sigma}_j^-$. This Hamiltonian emerges (without the factor of $2$) when one works in the single-excitation manifold and adiabatically eliminates the single-photon states~\cite{ref_pcw,ref_cirac_many-body, ref_jugal_paper-0}. $\hat{H}_{\text{single}}$ differs from $\hat{\Tilde{H}}_{\text{single}}$ because of the presence of the ``spectator", i.e., the constraint makes the Hamiltonian $\hat{H}_{\text{single}}$ considered in our work unique. To highlight the differences, red and blue circles in Fig.~\ref{fig_energy} show the eigenenergies of $\hat{H}_{\text{single}}$ and $\hat{\Tilde{H}}_{\text{single}}$, respectively, for $N_e=60$, $g/J=1/50$, $U/J=-1$, $\delta/J=-1/50$, and $x/a=1$. The constraint introduces an upshift of the eigenenergies for all eigenstates. The upshift is larger for the more negative eigenenergies (measured relative to the bottom $E_{0,b}$ of the two-photon bound state band) than the less negative eigenenergies. Interestingly, both $\hat{H}_{\text{single}}$ and $\hat{\Tilde{H}}_{\text{single}}$ support step like pattern, with each plateau containing close to $N_e$ eigenstates for the energetically lowest lying states. For the higher excited states, the steps are less pronounced. 
Reference~\cite{ref_rabl_atom-field} referred to the energy band formed by the qubit dominated states as a metaband.
While we observe, similarly to Ref.~\cite{ref_rabl_atom-field}, that the width of the band decreases with increasing $x$, it is important to point out that that work considered qubit-array physics in the single-excitation manifold on resonance (and not off-resonance as in our case) and for significantly stronger coupling strengths ($g/J$ of order $1$).

For comparison, the black circles in Fig.~\ref{fig_energy} show the eigenenergies for $\hat{H}_{\text{spin}}$. It can be seen that $\hat{H}_{\text{pair}}$ appreciably impacts the $10$ or so energetically lowest lying states and less so the higher-lying states. Importantly, the energies of the lowest few eigenstates of $\hat{H}_{\text{spin}}$ are pushed down due to the presence of the $\hat{H}_{\text{pair}}$ term. The downshift of the energies is associated with significant changes of the character of the eigenstates, i.e., a change from delocalized scattering states to localized bound states. We refer to this as ``pinning'' (see below for details). The energy spectrum shown in Fig.~\ref{fig_energy} is unique to a qubit spacing of $x/a=1$. For larger spacings, but otherwise identical parameters, the hopping energies are smaller and the step-like pattern is washed out. Moreover, the most strongly bound states are less separated from the other states than for $x/a=1$ (i.e., $\hat{H}_{\text{pair}}$ introduces a smaller downshift for the ground state for $x/a=2$ than for $x/a=1$). For $x=0$, there exist three degenerate 
energy levels: 
for the same $N_e$, $g/J$, $U/J$, and $\delta/J$ as considered in Fig.~\ref{fig_energy}, the $x/a=0$ spectrum for $\hat{H}_{\text{spin}}$ contains a single state with energy $E-E_{0,b}=-0.2256 J$, $N_e-1$ states with energy $E-E_{0,b}=-0.0629J$, and $N_e(N_e-3)/2$ states with energy $E-E_{0,b}=\delta=-0.02J$.

\begin{figure}[h]
\begin{center}
\includegraphics[width=0.35\textwidth]{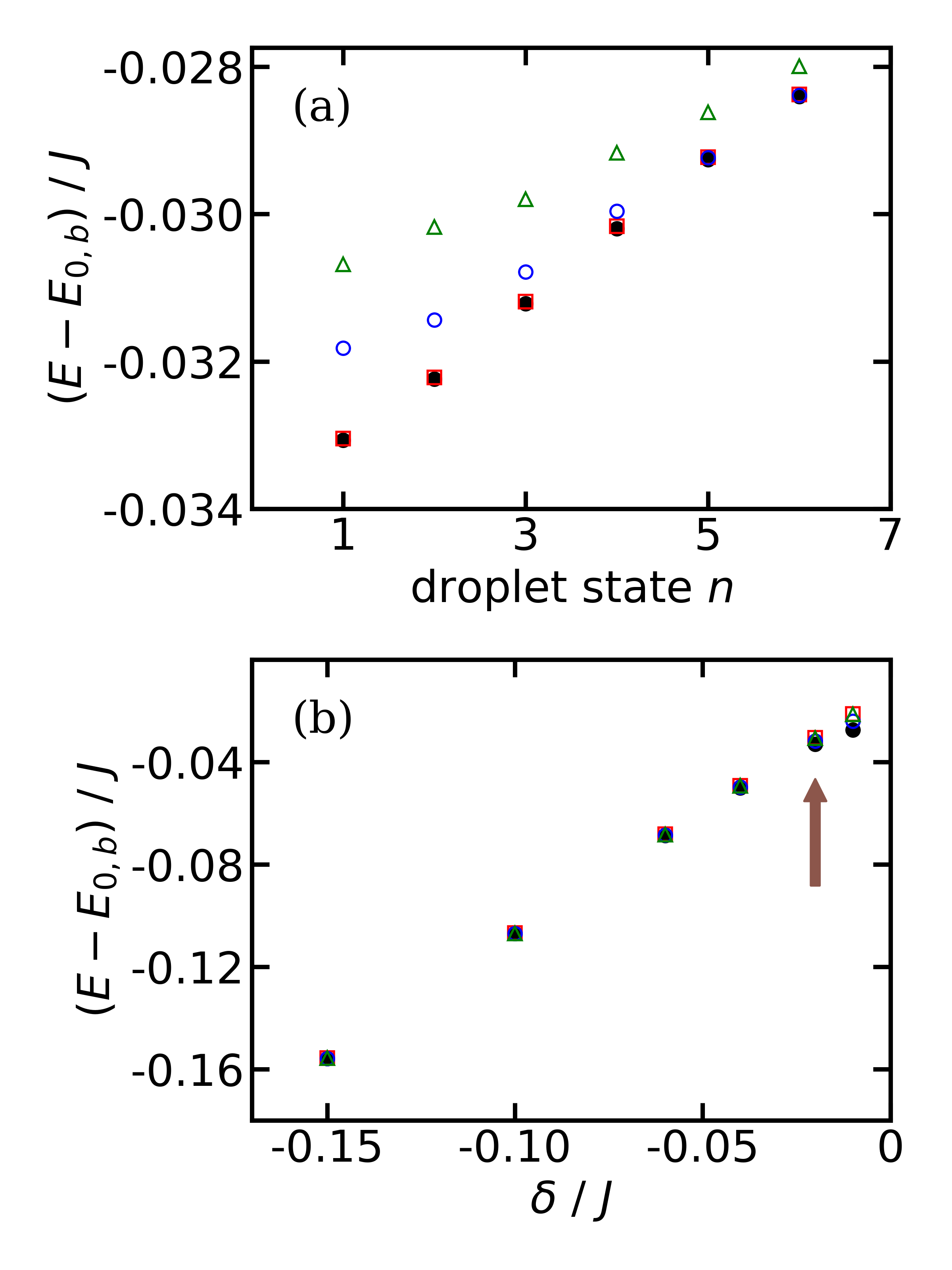}
\end{center}
\caption{Energy of droplet-like states, measured with respect to $E_{0,b}$, for $N_e=60$, $g/J=1/50$, $U/J=-1$, and $x/a=1$ as a function of (a) the excitation number $n$ for $\delta/J=-1/50$ and (b) the detuning $\delta/J$ for $n=1$. The black filled circles, green open triangles, red open squares, and blue open circles show the energies for $\hat{H}_{\text{spin}}$, $\hat{H}_{\text{single}}$, the variational wavefunction given in Eqs.~(\ref{eq_droplet-ansatz})-(\ref{eq_droplet-ansatz-q}), and for the perturbative calculation ($\hat{H}_{\text{pair}}$ is treated in first-order perturbation theory), respectively. In (a), the red and black symbols are nearly indistinguishable. In (b), all four calculations yield, on the scale shown, nearly indistinguishable energies except when $|\delta/J|$ is extremely small. The arrow in (b) marks the detuning used in (a).}
\label{fig_variational-perturbation}
\end{figure}

To understand the influence of $\hat{H}_{\text{pair}}$ on the eigenspectrum, we use first-order non-degenerate perturbation theory. Treating $\hat{H}_{\text{pair}}$ as  a perturbation, the first-order correction $E^{(1)}_n$ to the eigenenergy $E^{(0)}_n$ of the $n$th 
droplet-like eigenket
$\ket{\phi_n^{(0)}}$ of $\hat{H}_{\text{single}}$ is given by
\begin{eqnarray}
E^{(1)}_n=\bra{\phi^{(0)}_n}\hat{H}_{\text{pair}}\ket{\phi^{(0)}_n}.
\end{eqnarray}
Figure~\ref{fig_variational-perturbation}(a) considers the six energetically lowest-lying droplet-like states ($n=1-6$). These droplet-like states correspond to state numbers $1$, $2$, $3$, $4$, $7$, and $10$.
Figure~\ref{fig_variational-perturbation}(a) shows that the perturbation energies (the open blue circles show $E_n^{(0)}+E_n^{(1)}$) lie below the zeroth-order energies $E_n^{(0)}$ (green open triangles), 
i.e., the stronger binding of $\hat{H}_{\text{spin}}$ compared to $\hat{H}_{\text{single}}$ due to $\hat{H}_{\text{pair}}$ is captured qualitatively in first-order perturbation theory. Higher-order corrections, which  account for the mixing of the unperturbed states $\ket{\phi^{(0)}_n}$ play a larger role for the ground state ($n=1$) than for the excited droplet-like states ($n=2-6$). Figure~\ref{fig_variational-perturbation}(b) focuses on the energy of the lowest-lying droplet-like state and shows  
that
energy as a function of the detuning. The detuning marked by an arrow is identical to the detuning used in Fig.~\ref{fig_variational-perturbation}(a). For large to moderate detunings, the results from the perturbation calculation (blue open circles) agree well with the exact diagonalization of $\hat{H}_{\text{spin}}$ (black solid circles). For relatively small detunings, however, 
deviations are visible. While the first-order perturbative energy 
improves upon 
the
unperturbed energy, higher-order corrections play an increasingly more important role.

To characterize the eigenstates $\ket{\phi_E}$ of $\hat{H}_{\text{spin}}$, we expand them in terms of the basis 
kets
$\sigma^+_i\sigma^+_j\ket{g,\cdots,g}$,
 \begin{equation}
\label{eq_drop_coeff}
    \ket{\phi_E}=\sum_{i=1}^{N_e-1}\sum_{j=i+1}^{N_e}c^{(E)}_{i,j}\sigma^+_i\sigma^+_j\ket{g,\cdots,g},
\end{equation}
and analyze the expansion coefficients $c^{(E)}_{i,j}$ as well as the pair correlation function $P_{\text{pair}}(\alpha)$, which measures the likelihood that the two excitations are located at qubits that are separated by $\alpha$. The corresponding operator is given by
\begin{equation}
    \hat{P}_{\text{pair}}(\alpha)=\sum_{i=1}^{N_e-\alpha}\hat{\sigma}^+_i\hat{\sigma}^+_{i+\alpha}\ket{g,\cdots,g}\bra{g,\cdots,g}\hat{\sigma}^-_i\hat{\sigma}^-_{i+\alpha},
\end{equation}
where $\alpha$ takes the values $1$, $2$, $\cdots$, $N_e-1$. For example, if $\alpha=1$, the excitations are located at neighboring spins. In terms of the expansion coefficients, the pair correlation function for the eigenstate $\ket{\phi_E}$ is given by $P_{\text{pair}}(\alpha)=\sum_{i=1}^{N_e-\alpha}|c^{(E)}_{i,i+\alpha}|^2$.

\begin{figure}[h]
\includegraphics[width=0.35\textwidth]{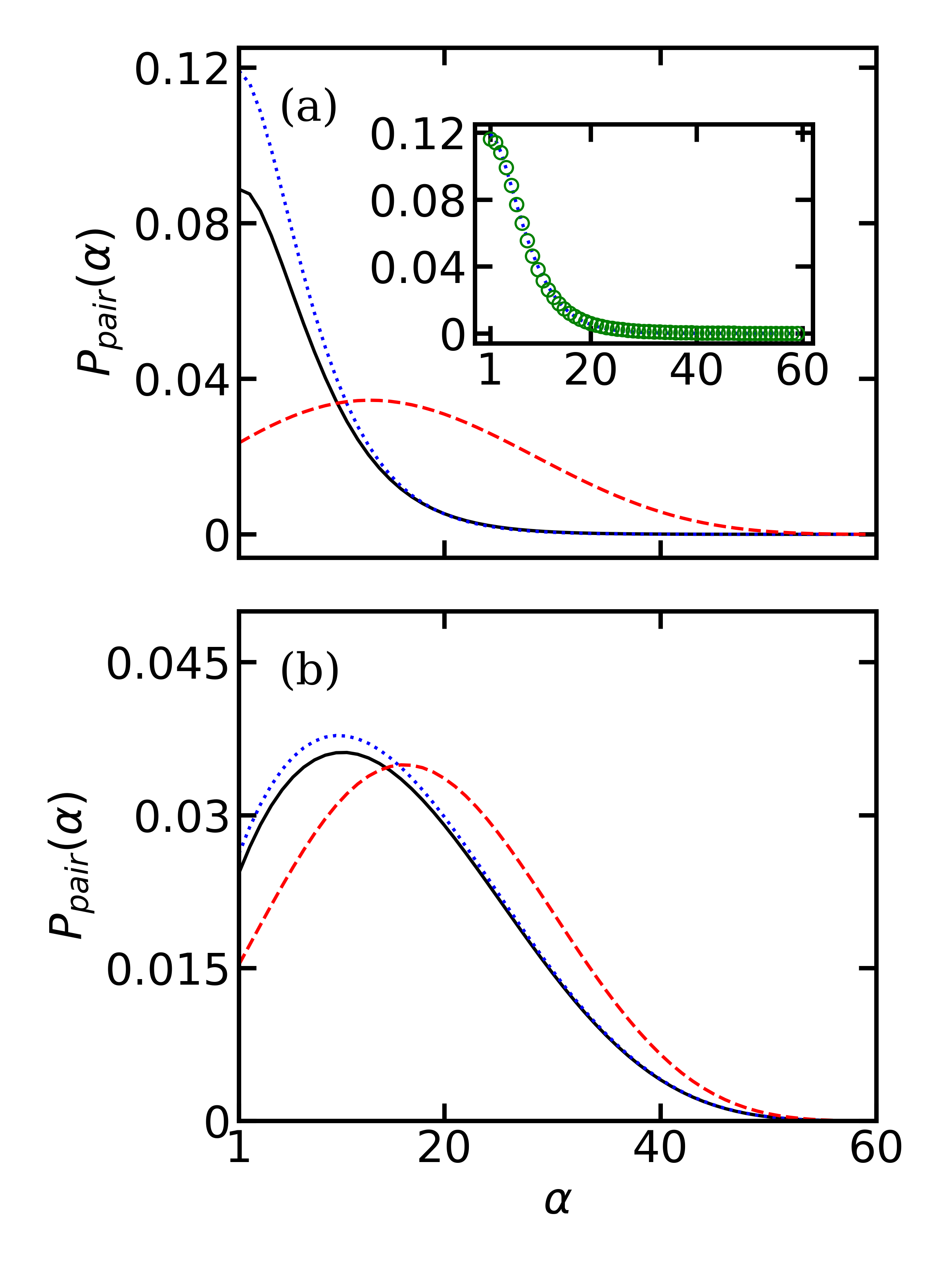}
\caption{Pair correlation function $P_{\text{pair}}(\alpha)$ for the ground state as a function of the separation $\alpha$ between two excited qubits for $N_e=60$, $g/J=1/50$, $U/J=-1$, $x/a=1$, and (a) $\delta/J=-1/50$ and (b) $\delta/J=-3/20$. The black solid, blue dotted, and red dashed lines are for $\hat{H}_{\text{full}}$, $\hat{H}_{\text{spin}}$, and $\hat{H}_{\text{single}}$, respectively. The inset in (a) replots the blue dotted line and additionally shows the variational results by green open circles.}
\label{fig_pair-correlation}
\end{figure}   

Figures~\ref{fig_pair-correlation}(a) and ~\ref{fig_pair-correlation}(b) show $P_{\text{pair}}(\alpha)$ for the ground state for $N_e=60$, $g/J=1/50$, $U/J=-1$, and $x/a=1$ for two different detunings, namely $\delta/J=-1/50$ and $-3/20$. The blue dotted lines are obtained using $\hat{H}_{\text{spin}}$. The full Hamiltonian $\hat{H}_{\text{full}}$ (black solid lines) yields results that are quite similar to those for $\hat{H}_{\text{spin}}$, thus providing  evidence that $\hat{H}_{\text{spin}}$ yields faithful results. For small $|\delta/J|$ [Fig.~\ref{fig_pair-correlation}(a)], the pair correlation function peaks at $\alpha=1$ and is essentially zero for $\alpha\gg 1$. This indicates that the two excited qubits want to stay together. The fall-off of $P_{\text{pair}}(\alpha)$ suggests that the ground state corresponds to a bound state. This interpretation is confirmed by calculations for larger arrays (larger $N_e$) with otherwise identical parameters. We find that $P_{\text{pair}}(\alpha)$ for the ground state remains essentially unchanged when $N_e$ is increased, i.e., the size of the ground state is independent of 
$N_e$, 
thereby justifying the classification as a self-bound state. 
For larger $|\delta/J|$ [Fig.~\ref{fig_pair-correlation}(b)], in contrast, the pair correlation function peaks at $\alpha\approx 10$ for $\hat{H}_{\text{full}}$ and $\hat{H}_{\text{spin}}$. This indicates that the two excited qubits have a tendency to spread out over the entire array. This interpretation is supported by the fact that the fall-off of the pair correlation function moves to larger $\alpha$ for larger $N_e$ but otherwise identical parameters. Correspondingly, we classify the ground state considered in Fig.~\ref{fig_pair-correlation}(b) as unbound. The inclusion of $\hat{H}_{\text{pair}}$ in the effective spin Hamiltonian $\hat{H}_{\text{spin}}$ (blue dotted line) is crucial. A comparison of the blue dotted line [$P_{\text{pair}}(\alpha)$ for $\hat{H}_{\text{spin}}$] and red dashed line [$P_{\text{pair}}(\alpha)$ for $\hat{H}_{\text{single}}$] reveals that $\hat{H}_{\text{pair}}$ has a pinning effect: it enhances, as already alluded to in Sec.~\ref{sec_approx_spin}, the probability to find excitations located at qubits that are close to each other. The effect is very prominent in Fig.~\ref{fig_pair-correlation}(a), where the red line is much broader than the blue line.
If  $\hat{H}_{\text{pair}}$ is neglected and $N_e$ is increased, the red line in Fig.~\ref{fig_pair-correlation} does not maintain its size, as is the case for $\hat{H}_{\text{spin}}$, but increases. This unequivocally shows that $\hat{H}_{\text{pair}}$  is responsible for the emergence of self-bound states.

\begin{figure}[h]
\includegraphics[width=0.47\textwidth]{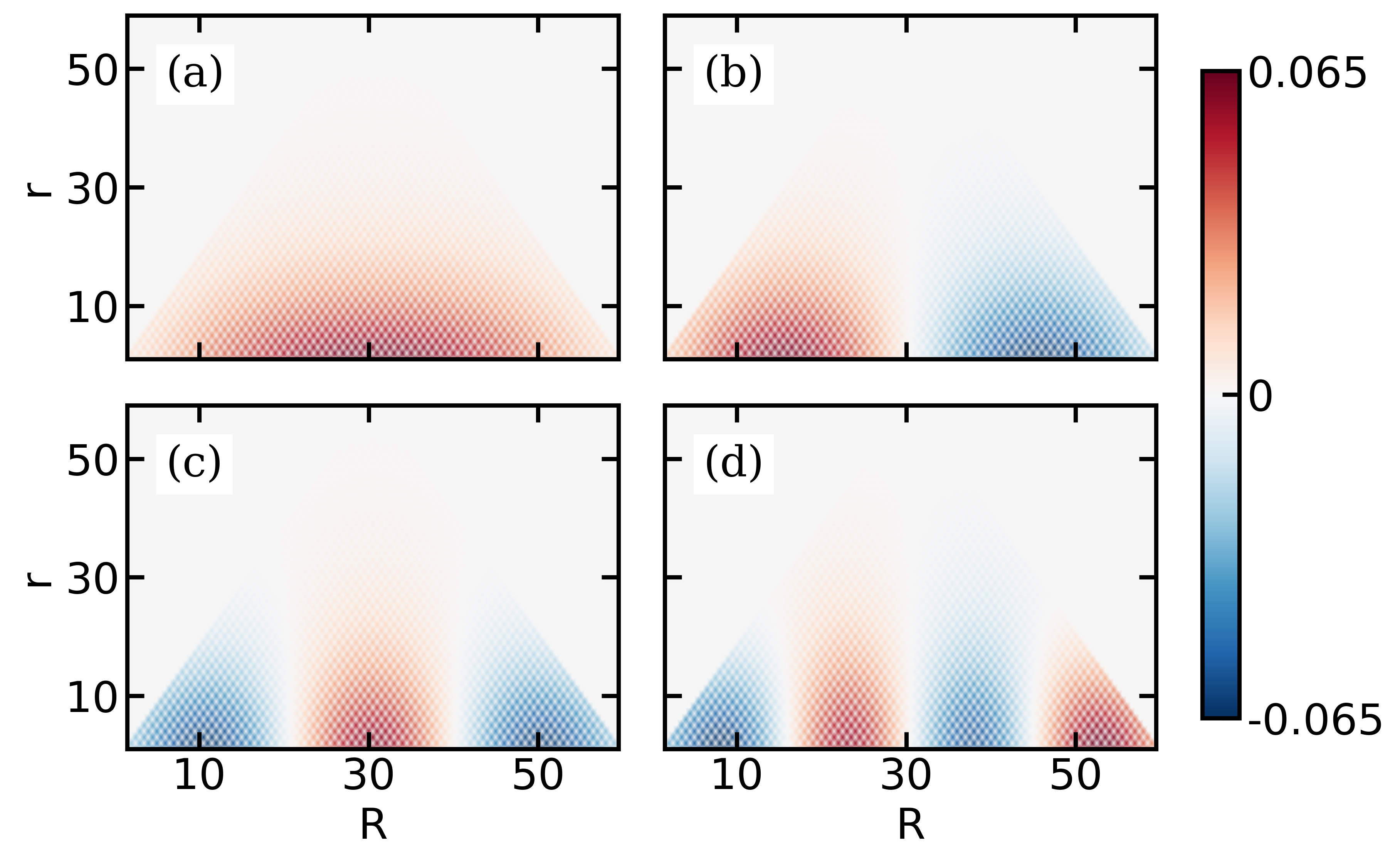}
\caption{Contour plots of the expansion coefficients $c^{(n)}_{ij}$ as  functions of $R$ and $r$ for $N_e=60$, $g/J=1/50$, $U/J=-1$, $\delta/J=-1/50$, and $x/a=1$. The coefficients are obtained by diagonalizing the effective Hamiltonian $\hat{H}_{\text{spin}}$. (a), (b), (c), and (d) are for $n$=$1$ 
(droplet-like
ground state), $2$ 
(droplet-like
first excited 
state), $3$ 
(droplet-like
second excited 
state),
and $4$ 
(droplet-like third excited 
state), respectively.}
\label{fig_droplets}
\end{figure}

Figure~\ref{fig_droplets} shows the real part of the coefficients $c_{i,j}^{(n)}$ for the four energetically lowest lying droplet-like bound states ($n=1-4$) for $N_e=60$, $g/J=1/50$, $U/J=-1$, $\delta/J=-1/50$, and $x/a=1$; the imaginary part is equal to zero. 
The droplet-like states shown in Fig.~\ref{fig_droplets} correspond to the state 
numbers
$1$, $2$, $3$, and $4$. 
Figure~\ref{fig_droplets} employs relative and center-of-mass coordinates $r$ and $R$, respectively, of the two excited qubits, $r=|j-i|$ and $R=(i+j)/2$.
The white area characterized by $r\geq 2R$ for $R<N_e/2$ and $r\geq 2(N_e-R)$ for $R\geq N_e/2$ is unphysical as there is a constraint of $i<j$ on the eigencoefficients due to the bosonic character or, equivalently, the exchange symmetry of the excitations. The small white dots, which exist in the physical $i<j$ portions in Fig.~\ref{fig_droplets}, result from the transformation from the ($i, j$) spin indices to the ($R, r$) coordinates. In Figs.~\ref{fig_droplets}(a)-\ref{fig_droplets}(d), the magnitude of the coefficients $c^{(n)}_{i,j}$ decreases with increasing $r$ for fixed $R$. Along the $R$ coordinate, the number of nodes increases from zero for the ground state [$n=1$ in Fig.~\ref{fig_droplets}(a)] to three for the third excited droplet-like state [$n=4$ in Fig.~\ref{fig_droplets}(d)]. The nodes are to a very good approximation parametrized by $R_{\text{node}}\approx \text{constant}$, i.e., they are, on the scale of Fig.~\ref{fig_droplets}, independent of $r$.

In what follows, we use a variational ansatz to understand the length scale that governs the droplet-like states and the number of droplet-like states that are supported by a qubit array of size $N_e$. Since Fig.~\ref{fig_droplets} suggests that the expansion coefficients of the $n$th droplet-like eigenstate decouple when plotted as functions of the relative coordinate $r$ and the center-of-mass coordinate $R$, we introduce the product ansatz
\begin{equation}
\label{eq_droplet-ansatz}
    c^{(n)}_{r,R}= Q^{(n)}(R)q(r).
\end{equation}
Here, the function $Q^{(n)}(R)$, 
\begin{equation}
\label{eq_droplet-ansatz-Q}
    Q^{(n)}(R)=\sqrt{\frac{2}{N_e}}\sin\left(\frac{n\pi}{N_e}R\right),
\end{equation} 
corresponds to the $n$th particle in the box wave function and the function $q(r)$,
\begin{equation}
\label{eq_droplet-ansatz-q}
 q(r)=2\sqrt{\frac{L_r^3}{\pi a^3}}\left[\frac{1}{(r-1)^2+(\frac{L_r}{a})^2}\right],   
\end{equation}
to an $n$-independent Lorentzian with characteristic length $L_r$. The length $L_r$ is treated as a variational parameter. By construction, the variational states with different $n$ are orthogonal. 

Figure~\ref{fig_variational-perturbation}(a) compares the variational energies (red open squares) of the six droplet-like states that are supported by the qubit array for $N_e=60$, $g/J=1/50$, $U/J=-1$, $\delta/J=-1/50$, and $x/a=1$ with those obtained by diagonalizing $\hat{H}_{\text{spin}}$ (black solid circles). We see that the variational energies agree extremely well with the exact eigenenergies of $\hat{H}_{\text{spin}}$. In Fig.~\ref{fig_variational-perturbation}(b), the energy of the ground droplet-like state is shown as a function of $\delta/J$ for the same $N_e$, $g/J$, $U/J$, and $x/a$ as used in Fig.~\ref{fig_variational-perturbation}(a). For large to moderate, in magnitude, detunings, the energies from the variational calculation (red open squares) agree well with the exact eigenenergies of $\hat{H}_{\text{spin}}$ (black solid circles). For small detunings, small deviations are visible. The variational calculation not only predicts the eigenenergy accurately but also the corresponding eigenstates. As an example, the green open circles in the inset of Fig.~\ref{fig_pair-correlation}(a) show the pair correlation function obtained by the variational treatment; it agrees well with the results obtained for $\hat{H}_{\text{spin}}$ (blue dotted line). The number of droplet-like states supported by the qubit array is approximately equal to $aN_e/L_r$. Intuitively, this can be understood as follows. The system develops additional nodes along the $R$ direction till the spacing between the nodes is comparable to the size of the droplet-like 
state
along the $r$ direction. For $g/J=1/50$, $U/J=-1$, $\delta/J=-1/50$, and $x/a=1$, the variational ground state energy is minimized for $L_r\approx 10a$. The qubit array with $N_e=60$ supports 
six
droplet-like states, in agreement with the estimate $aN_e/L_r\approx 6$. 
As the qubit array spacing $x$ is changed from 
$a$ to $2a$,
the number of droplet-like bound states decreases from six to four. For $x=3a$, droplet-like bound states are no longer supported.
Similar results are found for other parameter combinations.
We note that $\hat{H}_{\text{spin}}$ also supports more highly excited modes, which have nodes along the $r$-coordinate. The variational treatment of these energetically higher-lying droplet-like states is beyond the scope of this work.

\section{Dynamics}
\label{sec_dynamics}
This section discusses the dynamics for negative $\delta$ (band-gap regime) for two different initial states in the two-excitation manifold, namely the partially symmetric state $\ket{\text{PS}}$,
\begin{figure}[h]
\includegraphics[width=0.35\textwidth]{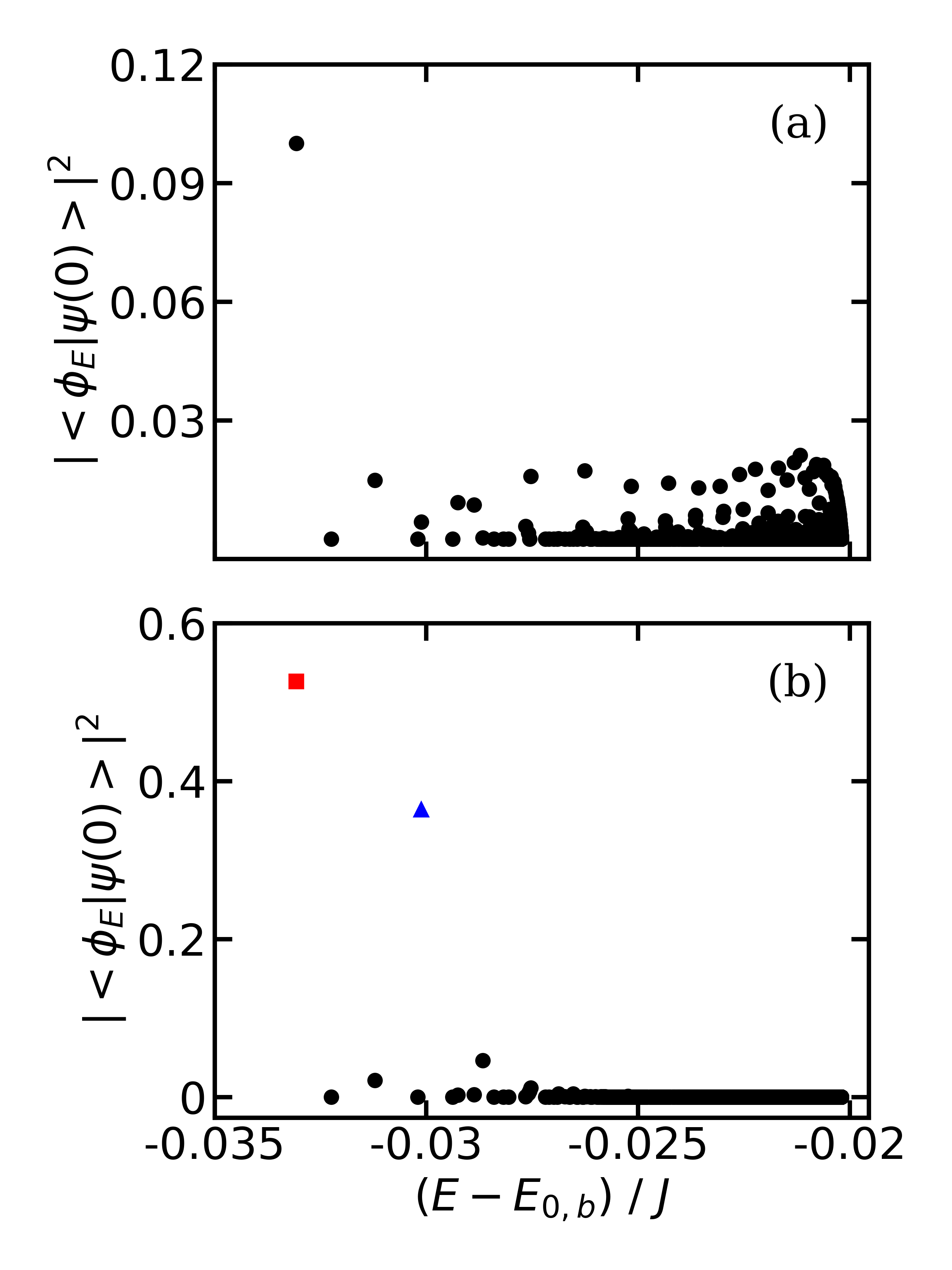}
\caption{Square of the absolute value of the projection of the initial state  $\ket{\psi(0)}$ onto the energy eigenstates $\ket{\phi_E}$ of $\hat{H}_{\text{spin}}$ as a function of the eigenenergy $E$, measured relative to the bottom $E_{0,b}$ of the two-photon bound state band, for $N_e=60$, $g/J=1/50$, $U/J=-1$, $\delta/J=-1/50$, and $x/a=1$. (a) The initial state is $\ket{\psi(0)}=\ket{\text{PS}}$. (b) The initial state is $\ket{\psi(0)}=\ket{\text{FS}}$. The red square and blue triangle correspond to the two largest values of $|\braket{\phi_E|\text{FS}}|^2$.}
\label{fig_pair-overlap}
\end{figure}   
\begin{equation}
    \ket{\text{PS}}=\frac{1}{\sqrt{N_e-1}}\sum_{i=1}^{N_e-1}\sigma_i^+\sigma_{i+1}^+\ket{g,\cdots,g},
\end{equation}
and the fully symmetric state $\ket{\text{FS}}$,
\begin{equation}
    \ket{\text{FS}}=\frac{\sqrt{2}}{\sqrt{N_e(N_e-1)}}\sum_{i=1}^{N_e-1}\sum_{j=i+1}^{N_e}\sigma_i^+\sigma_j^+\ket{g,\cdots,g}.
\end{equation}
The fully symmetric state is a superposition of all basis kets (all basis kets contribute with an expansion coefficient $\frac{\sqrt{2}}{\sqrt{N_e(N_e-1)}}$). The partially symmetric state, in contrast, only considers basis kets for which the excited qubits are nearest neighbors. 

Figures~\ref{fig_pair-overlap}(a) and~\ref{fig_pair-overlap}(b) show the decomposition of the states $\ket{\text{PS}}$ and $\ket{\text{FS}}$, respectively, into the energy eigenstates $\ket{\phi_E}$ of $\hat{H}_{\text{spin}}$ for $N_e=60$, $g/J=1/50$, $U/J=-1$, $\delta/J=-1/50$, and $x/a=1$. The state $\ket{\text{PS}}$ has finite overlap with a large number of eigenstates from all over the eigenspectrum. The ground state contributes about $10\%$ and the other states $3\%$ or less. For the state $\ket{\text{FS}}$ [Fig.~\ref{fig_pair-overlap}(b)], in contrast, there are two energy eigenstates that dominate and together contribute $89\%$ [$52.6\%$, red square in Fig.~\ref{fig_pair-overlap}(b), and $36.4\%$, blue triangle in Fig.~\ref{fig_pair-overlap}(b)]. The lowest eigenstate, which contributes $52.6\%$, has  
droplet-like
character while the excited eigenstate, which contributes $36.4\%$, has scattering 
characteristics. Since the fully symmetric initial state is dominated by two eigenstates, the dynamics is expected to feature Rabi-like two-state oscillation dynamics. The dynamics for the partially symmetric state, in contrast, is expected to display features of dephasing, at least over certain time scales, due to the 
superposition
of many energy eigenstates.

\begin{figure}[h]
\includegraphics[width=0.35\textwidth]{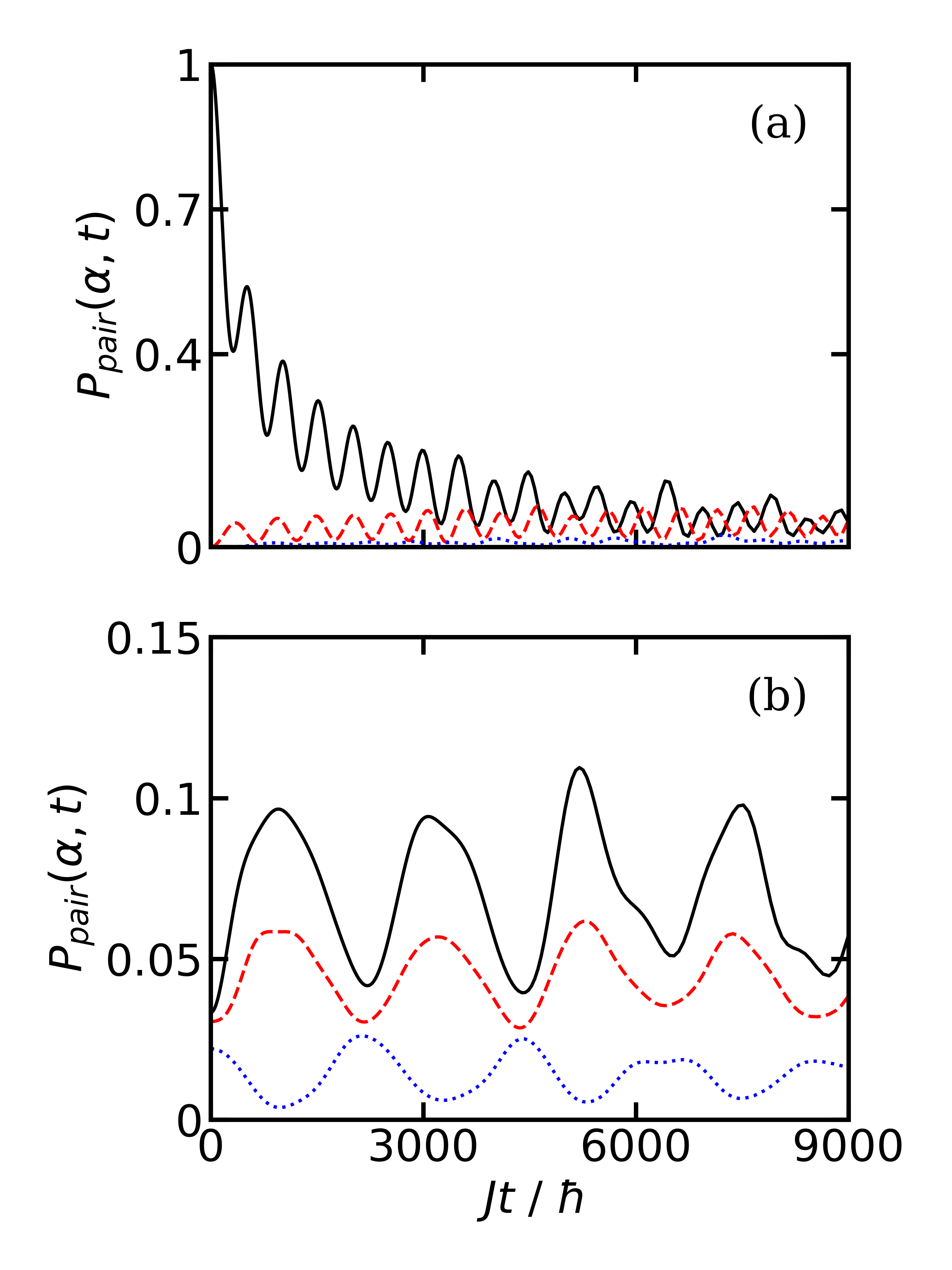}
\caption{Pair correlation function $P_{\text{pair}}(\alpha,t)$ for $\hat{H}_{\text{spin}}$ as a function of time for three different separations $\alpha$ for $N_e=60$, $g/J=1/50$, $U/J=-1$, $\delta/J=-1/50$, and $x/a=1$. The solid black, red dashed, and blue dotted lines show $P_{\text{pair}}(\alpha,t)$ for $\alpha=1$, $6$, and $21$, respectively. The initial state $\ket{\psi(0)}$ is equal to (a) $\ket{\text{PS}}$ and (b) $\ket{\text{FS}}$.}
\label{fig_oscillation}
\end{figure}  

Figure~\ref{fig_oscillation} shows the time dependence of the probability that two excitations belong to nearest neighbor qubits, i.e., qubits that are separated by 
$\alpha=1$ (black solid line), to qubits that are separated by $\alpha=6$ (red dashed line), and to qubits that are separated by $\alpha=21$ (blue dotted line). 
These observables are for the same parameters as those used in Fig.~\ref{fig_pair-overlap}. The time evolution of $P_{\text{pair}}(\alpha,t)$ for the initial states $\ket{\text{PS}}$ [Fig.~\ref{fig_oscillation}(a)] and $\ket{\text{FS}}$ [Fig.~\ref{fig_oscillation}(b)] is---as already anticipated based on the initial state decomposition---distinct. In Fig.~\ref{fig_oscillation}(a), $P_{\text{pair}}(\alpha,t)$ for $\alpha=1$ decays with damped oscillations. The damping or decay are attributed to the fact that a large number of eigenstates contribute to the initial state with comparable weight, giving rise to dephasing. In Fig.~\ref{fig_oscillation}(b), $P_{\text{pair}}(\alpha,t)$ oscillates with nearly undamped amplitude for all $\alpha$ considered. The slight ``distortions'' of the oscillations are caused by dephasing effects of the eigenstates that contribute to the fully symmetric initial state with a small weight, i.e., less than $5\%$. The oscillation period of $t\approx 2000\hbar/J$ corresponds to an energy of $0.0031J$. This energy agrees with the difference in energies of the two eigenstates that have the largest overlap with the initial state 
$\ket{\text{FS}}$ 
[red
square and blue triangle in Fig.~\ref{fig_pair-overlap}(b)].

\begin{figure}
\includegraphics[width=0.44\textwidth]{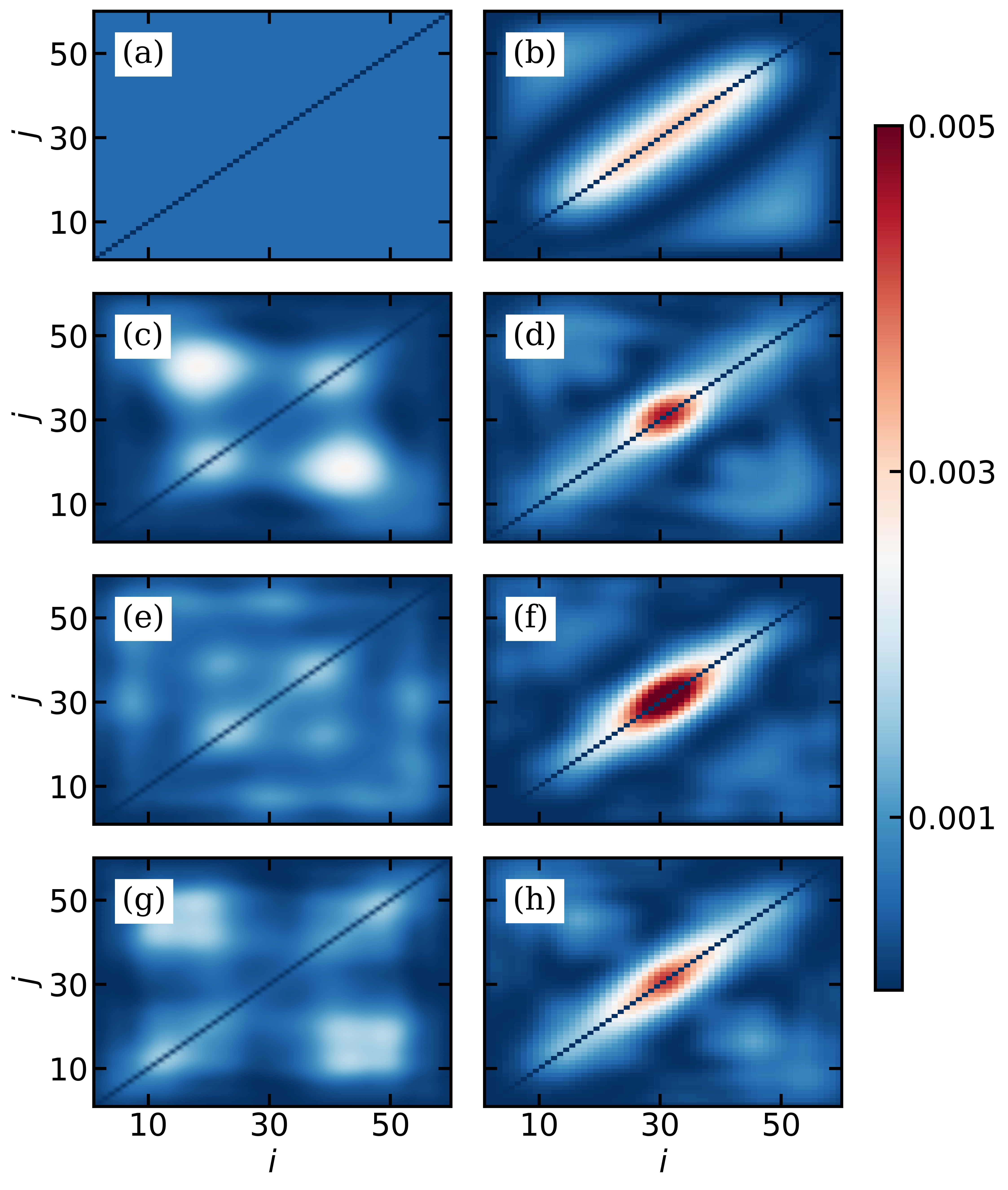}
\caption{Snapshots of $P_{\text{corr}}(i,j,t)$ for $\hat{H}_{\text{spin}}$ for $N_e=60$, $g/J=1/50$, $U/J=-1$, $\delta/J=-1/50$, and $x/a=1$ as functions of $i$ and $j$ for the initial state $\ket{\psi(0)}=\ket{\text{FS}}$. (a)-(h) correspond to $Jt/\hbar=0$, $960$, $2220$, $3080$, $4440$, $5220$, $6540$, and $7500$, respectively. As discussed in the main text in the context of Fig.~\ref{fig_droplets}, the coefficients $c_{i,j}^{(n)}$ are only defined  for $i<j$; to plot $P_{\text{corr}}(i,j,t)$, we artificially set $c_{i,j}^{(n)}=c_{j,i}^{(n)}$ for $i>j$ and $c_{i,j}^{(n)}=0$ for $i=j$ for ease of readibility.}
\label{fig_time-correlation}
\end{figure} 

Figure~\ref{fig_time-correlation} shows the spin-spin correlation function $P_{\text{corr}}(i,j,t)$, 
\begin{eqnarray}
\nonumber P_{\text{corr}}(i,j,t)=\bra{\psi(t)}\hat{\sigma}^+_i\hat{\sigma}^+_{j}\ket{g,\cdots,g} \times \\ \bra{g,\cdots,g}\hat{\sigma}^-_i\hat{\sigma}^-_{j}\ket{\psi(t)}
\end{eqnarray}
at eight different times ranging from zero in Fig.~\ref{fig_time-correlation}(a) to $Jt/\hbar=7500$ in Fig.~\ref{fig_time-correlation}(h) for $N_e=60$, $g/J=1/50$, $U/J=-1$, $\delta/J=-1/50$, and $x/a=1$. The initial state is $\ket{\text{FS}}$. The plots in the left column are for $Jt/\hbar=0$, $2220$, $4440$, and $6540$. Comparison with Fig.~\ref{fig_oscillation}(b) shows that $P_{\text{pair}}(\alpha=1,t)$ takes on a local minimum at these times. The plots in the right column of Fig.~\ref{fig_time-correlation}, in contrast, are such that $P_{\text{pair}}(\alpha=1,t)$ takes on a local maximum. We can see that $P_{\text{corr}}(i,j,t)$ is mostly concentrated around the middle of the diagonal in the right column while it is much more spread out in the left column. The observation that the spin-spin correlations alternate between being more localized and being more spread out can be readily explained by the fact that the initial state $\ket{\text{FS}}$ is dominated by contributions from the ground droplet-like state and a delocalized scattering state [red square and blue triangle Fig.~\ref{fig_pair-overlap}(b)]. This suggests that the droplet-like ground state can be probed by initializing the qubit array in the fully symmetric state $\ket{\text{FS}}$.

The calculations presented consider the ideal case scenario, where the excited state qubit has an infinite lifetime, the photon loss from the cavities is ignored, and imperfections---such as, e.g., a finite  spread $\Delta J$ of the tunneling energies $J$ and a finite spread $\Delta \omega_c$ of the cavity frequencies $\omega_c$ that exist to a varying degree in experiment---are neglected. To observe the oscillations displayed in Fig.~\ref{fig_oscillation}(b), the time scales associated with spontaneous qubit decay, photon losses, and ``dephasing" due to the spread of system parameters must be larger than about $10^4 \hbar/J$. In the following discussion, we assume that the excited state lifetime of the qubit is longer than the time scale for photon losses.

A finite ``bare" photon lifetime of $\hbar \kappa^{-1}$ leads to a characteristic decay time $(\Gamma_c)^{-1}$ that scales, for $|\delta_0| \ll 2J$, as
$\Gamma_c = p_{\text{ph}} \kappa / \hbar$, where $p_{\text{ph}}\approx g^2 J^{-1/2} \delta_0^{-3/2}/4$ and
$\delta_0$ denotes the detuning in the single-excitation manifold,   
$\delta_0 = (\hbar \omega_c - 2J) - \hbar \omega_e$~\cite{ref_rabl_non-linear,ref_rabl_atom-field}. Physically, the multiplicative factor $p_{\text{ph}}$ can be understood as arising from the admixture of the photonic degrees of freedom to the hybridized bound state in the single-excitation manifold.
Rewriting $\delta_0$ in terms of the detuning $\delta$ in the two-excitation manifold, we find
\begin{eqnarray}
p_{\text{ph}}= \frac{g^2}{4 \sqrt{J}} \left[ \frac{1}{2} \left( -\delta +\sqrt{U^2 + 16 J^2} \right)-2J \right]^{-3/2}.
\end{eqnarray}
For  $\delta/J=-1/50$ and $-3/20$, as used in this paper, $p_{\text{ph}}$ is equal to $5 \times 10^{-3}$ and $2 \times 10^{-3}$, respectively. 
To observe multiple oscillations, $(\Gamma_c)^{-1}$ must be much larger than $10^{4}\hbar/J$; the equal sign holds for $\kappa/J=2 \times 10^{-2}$ and $5 \times 10^{-2}$, respectively.
Superconducting circuit experiments have realized an eight cavity system with $U/h=-255$~MHz, $J/h=5-20$~MHz, and $\kappa/h=5$~kHz~\cite{ref_cqed-jon-simon}. This translates to
$\kappa/J=2.5 \times 10^{-4}$ to $10^{-3}$, i.e., experiments are already operating in a regime where the photon lifetime is sufficiently long to observe the predicted phenomena.
For fixed spreads $\Delta J$ and $\Delta \omega_c$, one may attempt to increase $\delta$ such that the spreads become, if measured as a multiple of the detuning, smaller. Since a larger $\delta$ corresponds to a smaller photon contribution $p_{\text{ph}}$ and hence a longer time scale for the photon losses,
there is some room to optimize the parameters for a specific experimental set-up.
While challenging, we conclude that the theory predictions put forward in this paper can be tested in state-of-the-art experiments.

\section{Conclusion}
\label{sec_conclusion}

This paper discussed the time-independent and time-dependent behaviors of a qubit array coupled to a non-linear photonic waveguide. Our interest was in the regime where the two-qubit transition energy lies in the band gap below the two-photon bound state band that is supported by the one-dimensional waveguide. We focused our attention on the two-excitation manifold. Even though the qubits are not interacting with each other, effective interactions---mediated by the waveguide---are introduced between qubits as a result of a two-step adiabatic elimination process. The resulting effective spin Hamiltonian, which was shown to accurately reproduce the key characteristics of the full Hamiltonian, features constrained single-qubit hopping and pair hopping interactions. The emergence of the latter critically depends on the presence of the Kerr-like non-linearity $U$. The effective spin Hamiltonian was shown to support a new class of  
droplet-like bound
states that arise due to the pair hopping interaction. These droplet-like states extend over many qubit lattice sites and can be probed dynamically. For the fully symmetric initial state, the populations were found to oscillate back and forth between a  
droplet-like 
bound 
state and a delocalized scattering state. While most of our discussion focused on $N_e=60$, $g/J=1/50$, $U/J=-1$, and $\delta/J=-1/50$, we emphasize that the characteristics discussed in this paper are also observed for other parameter 
combinations.

For fixed $g/J$, $\delta/J$, $N_e$, and $x/a$, we find that the number of droplet-like states supported by the qubit array decreases as $U/J$ becomes more negative. As $|U|/J$ increases, the two-photon bound state becomes more localized and hence the overall strength of 
the
pair hopping interaction becomes less negative. 
Whether
or not
droplet-like bound states exist also depends on the qubit array spacing $x$. If the separation between two neighboring qubits  
is increased, the number of droplet-like 
states
supported by the array decreases.

The giant droplet-like bound states discovered in this work provide an intriguing example of utilizing structured baths to engineer effective spin-spin interactions that support quantum states with non-trivial correlations. 
The droplet-like states considered in this paper, which emerge in the two-excitation sub-space, are distinct from the two-excitation scattering states considered in Ref.~\cite{ref_cirac_many-body} in the absence of the non-linearity $U$. They are also distinct from hydrid qubit-photon states that emerge in the single-excitation manifold~\cite{ref_rabl_non-linear}.
Possible extensions may focus on topological wave guides~\cite{ref_topo-1,ref_topo-2}, higher-dimensional baths, superlattice-type arrangements of the qubits, qubits with multiple transition frequencies~\cite{ref_scat-control}, multi-level emitters, qubits with multiple point contacts~\cite{ref_pohl-1}, and higher-excitation manifolds~\cite{ref_mult-1}. In all these scenarios, it will be interesting to explore the interplay between constrained single-qubit and pair hopping interactions.

\section{Acknowledgement}
\label{sec_acknowledgement}
Support by the National Science Foundation through grant number PHY-2110158 is gratefully acknowledged. This work used the OU Supercomputing Center for Education and Research (OSCER) at the University of Oklahoma (OU).

\appendix
\section{Derivation of $\hat{H}_{\text{spin}}$}
\label{append_A}
\begin{widetext}

\begin{figure}[h]
\includegraphics[width=0.90\textwidth]{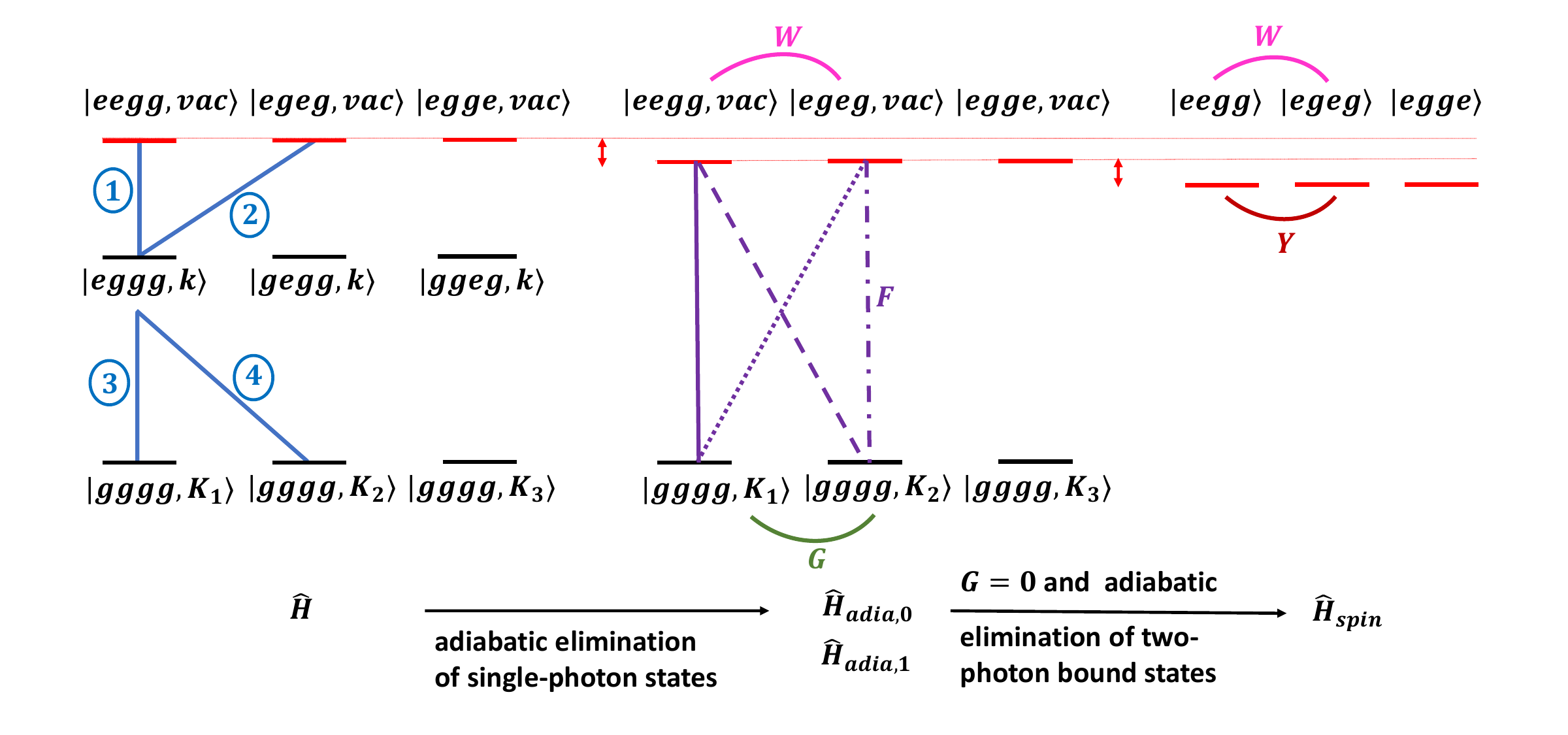}
\caption{Approximations made to obtain the effective spin Hamiltonian $\hat{H}_{\text{spin}}$ (right column) from the total Hamiltonian $\hat{H}$ (left column). The schematic considers $N_e=4$ as an example and shows only a subset of the basis kets. The red and black horizontal lines show a subset of basis kets. The blue, pink, purple, green, and dark red lines represent interactions. As a result of the adiabatic elimination of the single-photon states $\ket{eggg,k}$,  the interactions $1$ and $2$ give rise to the interaction $W$ between the states $\ket{eegg,\text{vac}}$ and $\ket{egeg,\text{vac}}$ (solid pink line), the interactions $1$ and $3$ give rise to the interaction $F$ between the states $\ket{eegg,\text{vac}}$ and $\ket{gggg,K_1}$ (solid purple line), the interactions $2$ and $3$ give rise to the interaction $F$ between the states $\ket{egeg,\text{vac}}$ and $\ket{gggg,K_1}$ (dotted purple line), the interactions $1$ and $4$ give rise to the interaction $F$ between the states $\ket{eegg,\text{vac}}$ and $\ket{gggg,K_2}$ (dashed purple line), the interactions $2$ and $4$ give rise to the interaction $F$ between the states $\ket{egeg,\text{vac}}$ and $\ket{gggg,K_2}$ (dash-dotted purple line), and the interactions $3$ and $4$ give rise to the interaction $G$ between the states $\ket{gggg,K_1}$ and $\ket{gggg,K_2}$ (green line). As a result of setting $G$ to zero and adiabatically eliminating the two-photon bound states $\ket{gggg,K}$, the interactions $F$ (e.g., solid and dotted lines, and dashed and dash-dotted lines) give rise to the interaction $Y$ between the states $\ket{eegg}$ and $\ket{egeg}$ (solid dark red line). The down shifts of the red basis states in the middle and right columns represent the  
energy 
shifts (sometimes called Stark shifts) that are due to the adiabatic eliminations.}
\label{fig_approx-scheme}
\end{figure}

\end{widetext}

Starting with the full Hamiltonian $\hat{H}$, this appendix derives the effective spin Hamiltonian $\hat{H}_{\text{spin}}$. The adiabatic elimination procedure discussed in this appendix is illustrated in Fig.~\protect\ref{fig_approx-scheme}.

\textit{Time-dependent wave packet}: Throughout, we assume that the two-photon scattering states can be neglected. This is justified since we are working in a regime where the two-qubit transition energy is far detuned from the two-photon scattering continuum. Under this approximation, the wavepacket $\ket{\psi(t)}$ in the two-excitation manifold can be written as~\cite{ref_rabl_non-linear, ref_jugal_paper-1, ref_jugal_paper-2} 
\begin{widetext}
\begin{eqnarray}
\label{eq_wavepacket_ansatz}
|\psi(t) \rangle  = \exp(- 2\imath \omega_e t)
\Bigg[
\sum_{i=1}^{N_e-1}\sum_{j=i+1}^{N_e}d_{ij}(t) \sigma^+_i \sigma^+_j|g,\cdots,g, \mbox{vac} \rangle + \sum_{i=1}^{N_e}\sum_k  c_{ik}(t) \sigma^+_i \hat{a}_k^{\dagger} |g,\cdots,g, \mbox{vac} \rangle + \nonumber \\ \sum_K c_{K,b}(t) \hat{B}_{K}^{\dagger} |g,\cdots,g, \mbox{vac}\rangle
\Bigg],
\end{eqnarray}
where $d_{ij}(t)$, $c_{ik}(t)$, and $c_{K,b}(t)$ denote expansion coefficients. 
The operator
$\hat{B}_{K}^{\dagger}$ 
creates 
a two-photon bound state with momentum $K$,
$|\psi_{K,b}\rangle=\hat{B}^{\dagger}|\text{vac}\rangle$. Inserting Eq.~(\ref{eq_wavepacket_ansatz}) into the time-dependent Schr\"odinger equation, we obtain a set of coupled differential equations 
\begin{eqnarray}
\label{eq_coeffe}
\imath \hbar \dot{d}_{ij}(t) = \frac{g}{\sqrt{N}}\sum_k \left[ \exp(\imath k a n_i) c_{j k}(t) + \exp(\imath k a n_j) c_{i k}(t)\right],
\end{eqnarray}
\begin{eqnarray}
\label{eq_coeffk}
\imath \hbar \dot{c}_{i k}(t) =
\Delta_k
c_{i k}(t)  
+\frac{g}{\sqrt{N}}\sum_{j=1,j\neq i}^{N_e} \exp( -\imath k a n_{j}) d_{\Tilde{i}\Tilde{j}}(t) +
\frac{g}{N} \sum_K M_b(k, n_{i},K) c_{K,b}(t),
\end{eqnarray}
and 
\begin{eqnarray}
\label{eq_coeffbound}
\imath \hbar \dot{c}_{K,b}(t) = 
\Delta_{K,b}
c_{K,b}(t) + \frac{g}{N} \sum_{i=1}^{N_e} \sum_k [M_b(k ,n_{i},K)]^* c_{i k}(t),
\end{eqnarray}
\end{widetext}
where $\Tilde{i}=\text{min}(i,j)$ and $\Tilde{j}=\text{max}(i,j)$. The energy detunings $\Delta_k$ and $\Delta_{K,b}$ are given by
\begin{eqnarray}
\Delta_k=E_k-\hbar \omega_e
\end{eqnarray}
and
\begin{eqnarray}
\Delta_{K,b}=E_{K,b}-2 \hbar \omega_e,
\end{eqnarray}
where $E_k$ denotes the energy of a single photon with wave vector $k$. The matrix elements $M_b(k, n,K)$ are defined as~\cite{ref_jugal_paper-1, ref_jugal_paper-2}
\begin{eqnarray}
\label{eq_mb_element}
\nonumber M_b(k, n,K)=\sqrt{2}\times \nonumber \\
\sum_m \exp \left[ \imath m \left(k-\frac{K}{2} \right)a 
+ \imath n (K-k)a \right] \psi_{K,b}(m),
\end{eqnarray}
where $\psi_{K,b}(m)=\langle ma|\psi_{K,b} \rangle$ is the two-photon bound state wave function ($ma$ denotes the relative distance between the two photons). Stationary and time-dependent solutions to the Schr\"{o}dinger equation for $\hat{H}_{\text{full}}$ are obtained through exact diagonalization, excluding the basis kets that span the two-photon scattering continuum. To characterize the distribution of the excited qubits, we monitor the pair correlation function $P_{\text{pair}}(\alpha,t)$,
\begin{eqnarray}
\nonumber P_{\text{pair}}(\alpha,t)=\bra{\psi(t)}\sum_{i=1}^{N_e-\alpha}\hat{\sigma}^+_i\hat{\sigma}^+_{i+\alpha}\ket{g,\cdots,g,\text{vac}}\times\\ 
\bra{g,\cdots,g,\text{vac}}\hat{\sigma}^-_i\hat{\sigma}^-_{i+\alpha}\ket{\psi(t)},
\end{eqnarray}
as well as the spin-spin correlation function $P_{\text{corr}}(i,j,t)$, 
\begin{eqnarray}
\nonumber P_{\text{corr}}(i,j,t)=\bra{\psi(t)}\hat{\sigma}^+_i\hat{\sigma}^+_{j}\ket{g,\cdots,g,\text{vac}}\times\\ \bra{g,\cdots,g,\text{vac}}\hat{\sigma}^-_i\hat{\sigma}^-_{j}\ket{\psi(t)}.     
\end{eqnarray}
In what follows, we introduce several approximations that eliminate the photonic degrees of freedom from the problem and, in turn, introduce effective interactions between groups of qubits.

\textit{Adiabatic elimination of the single-photon states:} Assuming that the changes of $c_{i k}(t)$ with time can be neglected, i.e., $\dot{c}_{i k}(t)=0$ in Eq.~(\ref{eq_coeffk}), the single-photon states $\hat{\sigma}^+_j\ket{g,\cdots,g,k}$ can be adiabatically eliminated. This approximation breaks down when $g$ is too large or the single-qubit transition energy is too close to the single-photon band. The resulting differential equations read
\begin{widetext}
\begin{eqnarray}
\label{eq_coeffeNew}
\imath \hbar \dot{d}_{ij}(t) = \sum_{l=1,l\neq j}^{N_e}W_{il}d_{\Tilde{l}\Tilde{j}}(t)+ \sum_{l=1,l\neq i}^{N_e}W_{lj}d_{\Tilde{i}\Tilde{l'}}(t) + \frac{g^2}{\sqrt{N}J}\sum_K F_{K,b}(n_i,n_j)c_{K,b}(t)
\end{eqnarray}
and
\begin{eqnarray}
\label{eq_coeffboundNew}
\imath \hbar \dot{c}_{K,b}(t) = \Delta_{K,b} c_{K,b}(t) +\frac{g^2}{\sqrt{N}J}\sum_{i=1}^{N_e}\sum_{j=i+1}^{N_e} F^*_{K,b}(n_i,n_j)d_{ij}(t)+\frac{g^2}{NJ}\sum_{K'}G_{KK'}(\Vec{n})c_{K',b}(t),
\end{eqnarray}
\end{widetext}
where $\Tilde{l}=\text{min}(l, j)$, $\Tilde{j}=\text{max}(l, j)$, $\Tilde{i}=\text{min}(l, i)$, $\Tilde{l'}=\text{max}(l, i)$, and $\Vec{n}=(n_1, n_2,\cdots, n_{N_e})$. It can be seen that the adiabatic elimination of the single-photon states introduces three effective interactions, namely $F_{K,b}$, $G_{KK'}$, and $W_{jl}$. The effective interaction $F_{K,b}(n_i,n_j)$ between the states $\hat{\sigma}_i^+\hat{\sigma}_j^+\ket{g,\cdots,g,\text{vac}}$ and $\hat{B}_K^{\dagger}\ket{g,\cdots,g,\text{vac}}$ is given by~\cite{ref_rabl_non-linear,ref_jugal_paper-1,ref_jugal_paper-2}
\begin{eqnarray}
\label{eq_fsubcapk}
\nonumber F_{K,b}(n_i,n_j)=\\
\nonumber -\sum_k\frac{J}{N\Delta_k}
\Big(\exp( -\imath k a n_i ) 
[M_b(k, n_j,K)]^*+\\
\exp(-\imath k a n_j) [M_b(k, n_i,K)]^*\Big).
\end{eqnarray}
The effective interaction $G_{KK'}(\Vec{n})$ between the states $\hat{B}_K^{\dagger}\ket{g,\cdots,g,\text{vac}}$ and  $\hat{B}_{K'}^{\dagger}\ket{g,\cdots,g,\text{vac}}$ is given by~\cite{ref_rabl_non-linear,ref_jugal_paper-1,ref_jugal_paper-2}
\begin{eqnarray}
\label{eq_GKKprime}
\nonumber G_{KK'}(\Vec{n})=-
 \sum_{j=1}^{N_e}\sum_k 
\frac{J}{N\Delta_k}
[M_b(k,n_{j},K)]^* \times \\M_b(k, n_{j},K').
\end{eqnarray}
The interactions $F_{K,b}$ and $G_{KK'}$ have been discussed extensively in the context of the two-qubit system ($\hat{H}$ with $N_e=2$)~\cite{ref_rabl_non-linear,ref_jugal_paper-1, ref_jugal_paper-2}. The effective interaction $W_{jl}$, in contrast, does not exist for $N_e=2$; it critically depends on having more than two qubits coupled to the cavity array. The functional form of $W_{jl}$ is given in Eq.~(\ref{eq_W}) of the main text. 

Equations~(\ref{eq_coeffeNew})--(\ref{eq_coeffboundNew}) correspond to the effective Hamiltonian $\hat{H}_{\text{adia,0}}$,
\begin{widetext}
\begin{eqnarray}
\label{eq_H_adia0}
 \nonumber \hat{H}_{\text{adia,0}} = \hat{H}_{\text{single}}+ \frac{g^2}{J\sqrt{N}}
 \sum_{i=1}^{N_e}\sum_{j=i+1}^{N_e}\sum_K\left[F_{K,b}(n_i,n_j) \hat{\sigma}^+_i\hat{\sigma}^+_j \hat{B}_K + F^*_{K,b}(n_i,n_j) \hat{\sigma}^-_i\hat{\sigma}^-_j \hat{B}^{\dagger}_K\right]+\\ \sum_K \Delta_K \hat{B}_K^{\dagger}\hat{B}_K + \frac{g^2}{NJ}\sum_K\sum_{K'}G_{KK'}(\Vec{n})\hat{B}_K^{\dagger}\hat{B}_{K'},
\end{eqnarray}
\end{widetext}
where $\hat{H}_{\text{single}}$ is given in Eq.~(\ref{eq_H-hopping}) of the main text. For $N_e=2$, Refs.~\cite{ref_jugal_paper-1,ref_jugal_paper-2} found that the effective interaction $G_{KK'}$ plays a non-negligible role only when the transition energy $2\hbar\omega_e$ of two qubits is in or nearly in resonance with the bottom of the two-photon bound state band. Since $G_{KK'}$ plays, in general, a negligible role away from the bottom of the band, it is useful to define the effective Hamiltonian $\hat{H}_{\text{adia,1}}$ by setting $G_{KK'}$ in $\hat{H}_{\text{adia,0}}$ to zero. The effective Hamiltonians $\hat{H}_{\text{adia,0}}$ and $\hat{H}_{\text{adia,1}}$ live in the $\left(\frac{N_e(N_e-1)}{2}+N\right)$--dimensional Hilbert space that is spanned by the states $\hat{\sigma}^+_i\hat{\sigma}^+_j\ket{g,\cdots,g, \text{vac}}$ and $\hat{B}^{\dagger}_K\ket{g,\cdots,g,\text{vac}}$ with wave vector $K$.

\textit{Adiabatic elimination of the two-photon bound states:} For the band gap physics considered in this paper, the energy $2\hbar\omega_e$ of two excited qubits is not in resonance with the two-photon bound state band. Consequently, we adiabatically eliminate the two-photon bound states, i.e., we set the left hand side of Eq.~(\ref{eq_coeffboundNew}) to zero. Using this to eliminate $c_{K,b}(t)$ from Eq.~(\ref{eq_coeffeNew}), the resulting set of coupled 
equations---setting $G_{KK'}=0$---reads
\begin{eqnarray}
\label{eq_coeffSpin}
\nonumber\imath \hbar \dot{d}_{ij}(t) = \sum_{l=1,l\neq j}^{N_e}W_{il}d_{\Tilde{l}\Tilde{j}}(t)+ \sum_{l=1,l\neq i}^{N_e}W_{lj}d_{\Tilde{i}\Tilde{l'}}(t) +\\
\sum_{l=1}^{N_e-1}\sum_{h=l+1}^{N_e}Y_{ij,lh}d_{lh}(t).
\end{eqnarray}
Equation~(\ref{eq_coeffSpin}) corresponds to the effective spin Hamiltonian $\hat{H}_{\text{spin}}$ given in Eq.~(\ref{eq_H_eff2}) of the main text, which lives in the $N_e(N_e-1)/2$--dimensional Hilbert spanned by the qubit states $\hat{\sigma}^+_i\hat{\sigma}^+_j\ket{g,\cdots,g}$.


\begin{thebibliography}{100}

\bibitem{ref_sinha}
K. Sinha, P. Meystre, E. A. Goldschmidt, F. K. Fatemi, S. L. Rolston, and P. Solano, Non-Markovian Collective Emission from Macroscopically Separated Emitters, Phys. Rev. Lett. 124, 043603 (2020).

\bibitem{ref_chiral-zoller}
H. Pichler, T. Ramos, A. J. Daley, and P. Zoller, Quantum optics of chiral spin networks, Phys. Rev. A 91, 042116 (2015).

\bibitem{ref_Haakh}
H. R. Haakh, S. Faez, and V. Sandoghdar, Polaritonic normal-mode splitting and light localization in a one-dimensional nanoguide, Phys. Rev. A 94, 053840 (2016).

\bibitem{ref_chaos}
A. V. Poshakinskiy, J. Zhong, and A. N. Poddubny, Quantum Chaos Driven by Long-Range Waveguide-Mediated Interactions, Phys. Rev. Lett. 126, 203602 (2021).

\bibitem{ref_henriet}
L. Henriet, Z. Ristivojevic, P. P. Orth, and K. L. Hur, Quantum dynamics of the driven and dissipative Rabi model, Phys. Rev. A 90, 023820 (2014).

\bibitem{ref_wolf}
F. A. Wolf, M. Kollar, and D. Braak, Exact real-time dynamics of the quantum Rabi model, Phys. Rev. A 85, 053817 (2012).

\bibitem{ref_hwang}
M.-J. Hwang, R. Puebla, and M. B. Plenio, Quantum Phase Transition and Universal Dynamics in the Rabi Model, Phys. Rev. Lett. 115, 180404 (2015).


\bibitem{ref_altintas}
F. Altintas and R. Eryigit, Dissipative dynamics of quantum correlations in the strong-coupling regime, Phys. Rev. A 87, 022124 (2013).

\bibitem{ref_mahmoodian}
S. Mahmoodian, Chiral Light-Matter Interaction beyond the Rotating-Wave Approximation, Phys. Rev. Lett. 123, 133603 (2019).

\bibitem{ref_kockum}
A. F. Kockum, A. Miranowicz, S. D. Liberato, S. Savasta, and F. Nori, Ultrastrong coupling between light and matter, Nat. Rev. Phys. 1, 19 (2019).

\bibitem{ref_diaz}
P. Forn-Díaz, L. Lamata, E. Rico, J. Kono, and E. Solano, Ultrastrong coupling regimes of light-matter interaction, Rev. Mod. Phys. 91, 025005 (2019).


\bibitem{ref_ripoll-1}
E. Sánchez-Burillo, D. Zueco, L. Martín-Moreno, and J. J. García-Ripoll, Dynamical signatures of bound states in waveguide QED, Phys. Rev. A 96, 023831 (2017).

\bibitem{ref_ripoll-2}
G. Díaz-Camacho, D. Porras, and J. J. García-Ripoll, Photon-mediated qubit interactions in one-dimensional discrete and continuous models, Phys. Rev. A 91, 063828 (2015).

\bibitem{ref_shi}
T. Shi, Y.-H. Wu, A. González-Tudela, and J. I. Cirac, Bound States in Boson Impurity Models, Phys. Rev. X 6, 021027 (2016).

\bibitem{ref_rabl_atom-field}
G. Calajó, F. Ciccarello, D. Chang, and P. Rabl, Atom-field dressed states in slow-light waveguide QED, Phys. Rev. A 93, 033833 (2016).


\bibitem{ref_pcw}
J. Douglas, H. Habibian, C.-L. Hung, A. V. Gorshkov, H. J. Kimble, and D. E. Chang, Quantum many-body models with cold atoms coupled to photonic crystals, Nature Photon. 9, 326–331 (2015). 

\bibitem{ref_cirac_many-body}
T. Shi, Y-H Wu, A. González-Tudela, and J. I. Cirac, Effective many-body Hamiltonians of qubit-photon bound states, New J. Phys. 20, 105005 (2018).

\bibitem{ref_jugal_paper-0}
Y. Chougale, J. Talukdar, T. Ramos, and R. Nath, Dynamics of Rydberg excitations and quantum correlations in an atomic array coupled to a photonic crystal waveguide, Phys. Rev. A 102, 022816 (2020).

\bibitem{ref_google-group}
A. Morvan et al., Formation of robust bound states of interacting microwave photons, Nature 612, 240–245 (2022).


\bibitem{ref_gorshkov_ph-ph}
A. V. Gorshkov, J. Otterbach, M. Fleischhauer, T. Pohl, and M. D. Lukin, Photon-Photon Interactions via Rydberg Blockade, Phys. Rev. Lett. 107, 133602 (2011).

\bibitem{ref_jeanic_ph-ph}
H. L. Jeannic, A. Tiranov, J. Carolan, T. Ramos, Y. Wang, M. H. Appel, S. Scholz, A. D. Wieck, A. Ludwig, N. Rotenberg, L. Midolo, J. J. García-Ripoll, A. S. Sørensen, and P. Lodahl, Dynamical photon–photon interaction mediated by a quantum emitter, Nat. Phys. 18, 1191–1195 (2022).

\bibitem{ref_pohl_ph-ph}
L. Zhang, V. Walther, K. Mølmer, and T. Pohl, Photon-
photon interactions in Rydberg-atom arrays, Quantum
6, 674 (2022).

\bibitem{ref_roy}
D. Roy, C. M. Wilson, and O. Firstenberg, Colloquium: Strongly interacting photons in one-dimensional continuum, Rev. Mod. Phys. 89, 021001 (2017).

\bibitem{ref_nano-shah}
E. Shahmoon, P. Grišins, H. P. Stimming, I. Mazets, and G. Kurizki, Highly nonlocal optical nonlinearities in atoms trapped near a waveguide, Optica 3, 725-733 (2016). 

\bibitem{ref_palma}
F. Lombardo, F. Ciccarello, and G. M. Palma, Photon localization versus population trapping in a coupled-cavity array, Phys. Rev. A 89, 053826 (2014).

\bibitem{ref_few-ph}
P. Longo, P. Schmitteckert, and K. Busch, Few-Photon Transport in Low-Dimensional Systems: Interaction-Induced Radiation Trapping, Phys. Rev. Lett. 104, 023602 (2010).

\bibitem{ref_wg_many-bound}
H. Zheng, D. J. Gauthier, and H. U. Baranger, Waveguide QED: Many-body bound-state effects in coherent and Fock-state scattering from a two-level system, Phys. Rev. A 82, 063816 (2010).


\bibitem{ref_reitz}
D. Reitz, C. Sayrin, R. Mitsch, P. Schneeweiss, and A. Rauschenbeutel, Coherence Properties of Nanofiber-Trapped Cesium Atoms, Phys. Rev. Lett. 110, 243603 (2013).

\bibitem{ref_yalla}
R. Yalla, M. Sadgrove, K. P. Nayak, and K. Hakuta, Cavity Quantum Electrodynamics on a Nanofiber Using a Composite Photonic Crystal Cavity, Phys. Rev. Lett. 113, 143601 (2014).

\bibitem{ref_hood}
J. D. Hood, A. Goban, A. Asenjo-Garcia, M. Lu, S.-P. Yu, D. E. Chang, and H. J. Kimble, Atom-atom interactions around the band edge of a photonic crystal waveguide, Proc. Natl. Acad. Sci. U.S.A. 113, 10507 (2016).

\bibitem{ref_resonator}
L. Zhou, Z. R. Gong, Y.-X. Liu, C. P. Sun, and F. Nori, Controllable Scattering of a Single Photon inside a One-Dimensional Resonator Waveguide, Phys. Rev. Lett. 101, 100501 (2008).

\bibitem{ref_sqbit}
J.-T. Shen and S. Fan, Coherent Single Photon Transport in a One-Dimensional Waveguide Coupled with Superconducting Quantum Bits, Phys. Rev. Lett. 95, 213001 (2005). 

\bibitem{ref_photon-mat_cqed}
I. Carusotto, A. A. Houck, A. J. Kollár, P. Roushan, D. I. Schuster, and J. Simon, Photonic materials in circuit quantum electrodynamics, Nat. Phys. 16, 268–279 (2020).


\bibitem{ref_cqed-jon-simon}
R. Ma, B. Saxberg, C. Owens, N. Leung, Y. Lu, J. Simon, and D. I. Schuster, A dissipatively stabilized Mott insulator of photons, Nature 566, 51–57 (2019).

\bibitem{ref_plasmom}
D. Chang, A. S. Sørensen, E. Demler, and M. D. Lukin, A single-photon transistor using nanoscale surface plasmons, Nat. Phys. 3, 807–812 (2007).

\bibitem{ref_saffman}
M. Saffman, T. G. Walker, and K. Mølmer, Quantum information with Rydberg atoms, Rev. Mod. Phys. 82, 2313 (2010).

\bibitem{ref_levine}
H. Levine, A. Keesling, A. Omran, H. Bernien, S. Schwartz, A. S. Zibrov, M. Endres, M. Greiner, V. Vuletić, and M. D. Lukin, High-Fidelity Control and Entanglement of Rydberg-Atom Qubits, Phys. Rev. Lett. 121, 123603 (2018).


\bibitem{ref_qdot}
P. Lodahl, S. Mahmoodian, and S. Stobbe, Interfacing
single photons and single quantum dots with photonic
nanostructures, Rev. Mod. Phys. 87, 347 (2015).


\bibitem{ref_astafiev}
O. Astafiev, A. M. Zagoskin, A. A. Abdumalikov, Yu. A.
Pashkin, T. Yamamoto, K. Inomata, Y. Nakamura, and J. S. Tsai, Resonance fluorescence of a single artificial atom, Science 327, 840 (2010).

\bibitem{ref_hoi}
I.-C. Hoi, C. M. Wilson, G. Johansson, T. Palomaki, B. Peropadre, and P. Delsing, Demonstration of a Single- Photon Router in the Microwave Regime, Phys. Rev. Lett.
107, 073601 (2011).

\bibitem{ref_mlynek}
J. A. Mlynek, A. A. Abdumalikov, C. Eichler, and A.
Wallraff, Observation of Dicke superradiance for two artificial atoms in a cavity with high decay rate, Nat.
Commun. 5, 5186 (2014).

\bibitem{ref_mirhosseini}
M. Mirhosseini, E. Kim, X. Zhang, A. Sipahigil, P. B.
Dieterle, A. J. Keller, A. Asenjo-Garcia, D. E. Chang, and
O. Painter, Cavity quantum electrodynamics with atom-like
mirrors, Nature (London) 569, 692 (2019).

\bibitem{ref_sundaresan}
N. M. Sundaresan, R. Lundgren, G. Zhu, A. V. Gorshkov,
and A. A. Houck, Interacting Qubit-Photon Bound States
with Superconducting Circuits, Phys. Rev. X 9, 011021
(2019).

\bibitem{casas1990}
M. Casas and S. Stringari,
Elementary Excitations of $^4$He Clusters, J. Low Phys. Temp. {\bf{79}}, 135 (1990).

\bibitem{ref_rabl_non-linear}
Z. Wang, T. Jaako, P. Kirton, and P. Rabl, Supercorrelated Radiance in Nonlinear Photonic Waveguides, Phys. Rev. Lett. 124, 213601 (2020).

\bibitem{ref_jugal_paper-1}
J. Talukdar and D. Blume, Undamped Rabi oscillations due to polaron-emitter hybrid states in a nonlinear photonic waveguide coupled to emitters, Phys. Rev. A 106, 013722 (2022).

\bibitem{ref_jugal_paper-2}
J. Talukdar and D. Blume, Two emitters coupled to a bath with Kerr-like nonlinearity: Exponential decay, fractional populations, and Rabi oscillations, Phys. Rev. A 105, 063501 (2022).

\bibitem{ref_john}
S. John and J. Wang, Quantum electrodynamics near a photonic band gap: Photon bound states and dressed atoms, Phys. Rev. Lett. 64, 2418 (1990).


\bibitem{ref_XXX-bethe}
J. M\"{o}lter, T. Barthel, U. Schollw\"{o}ck, and V. Alba, Bound states and entanglement in the excited states of quantum spin chains, J. Stat. Mech. 2014, P10029.

\bibitem{ref_wang-nature}
Z. Wang, J. Wu, W. Yang, A. K. Bera, D. Kamenskyi, A. T. M. N. Islam, S. Xu, J. M. Law, B. Lake, C. Wu, and A. Loidl, Experimental observation of bethe strings,  
Nature 554,
219-223 (2018).


\bibitem{ref_molmer1}
R. Piil and K. Mølmer, Tunneling couplings in discrete lattices, single-particle band structure, and eigenstates of interacting atom pairs, Phys. Rev. A 76, 023607 (2007).

\bibitem{ref_molmer2}
N. Nygaard, R. Piil, and K. Mølmer, Two-channel Feshbach physics in a structured continuum, Phys. Rev. A 78, 023617 (2008).

\bibitem{ref_valiente}
 M. Valiente and D. Petrosyan, Two-particle states in the Hubbard model, J. Phys. B: At. Mol. Opt. Phys. 41, 161002 (2008).
 
 \bibitem{ref_petrosyan}
 D. Petrosyan, B. Schmidt, J. R. Anglin, and M. Fleischhauer, Quantum liquid of repulsively bound pairs of particles in a lattice, Phys. Rev. A 76, 033606 (2007).

\bibitem{ref_atw-1}
S. J. Masson and A. Asenjo-Garcia, Atomic-waveguide quantum electrodynamics, Phys. Rev. Res. 2, 043213 (2020).

\bibitem{ref_atw-2}
D. Castells-Graells, D. Malz, C. C. Rusconi, and J. I. Cirac, Atomic waveguide QED with atomic dimers, Phys. Rev. A 104, 063707 (2021).

\bibitem{ref_topo-1}
M. Bello, G. Platero, J. I. Cirac, and A. Gonz{\'a}lez-Tudela, Unconventional Quantum Optics in Topological Waveguide QED, Sci. Adv. 5, eaaw0297 (2019).

\bibitem{ref_topo-2}
E. Kim, X. Zhang, V. S. Ferreira, J. Banker, J. K. Iverson, A. Sipahigil, M. Bello, A. González-Tudela, M. Mirhosseini, and O. Painter, Quantum Electrodynamics in a Topological Waveguide, Phys. Rev. X 11, 011015 (2021).

\bibitem{ref_scat-control}
L. Qiao and J. Gong, Coherent Control of Collective Spontaneous Emission through Self-Interference, Phys. Rev. Lett. 129, 093602 (2022).

\bibitem{ref_pohl-1}
O. A. Iversen and T. Pohl, Self-ordering of individual photons in waveguide QED and Rydberg-atom arrays, Phys. Rev. Res. 4, 023002 (2022).

\bibitem{ref_mult-1}
O. Rubies-Bigorda, S. Ostermann, and S. F. Yelin, Generating multi-excitation subradiant states in incoherently excited atomic arrays,  	arXiv:2209.00034.
 

\end{thebibliography}
\end{document}